\documentclass[useAMS,usenatbib]{mn2e}

\usepackage{graphicx,color}
\usepackage{textcomp}
\usepackage{amsmath,amssymb}
\usepackage{comment}

\DeclareGraphicsExtensions{.eps, .ps}

\renewcommand{\arcsec}{$^{\prime\prime}$}

\newcommand{\qm}[1]{``#1''}

\newcommand{\ha}{H$\alpha$}
\newcommand{\Hb}{H$\beta$}
\newcommand{\oiii}{[O\,{\scriptsize III}\,]}
\newcommand{\nii}{[N\,{\scriptsize II}\,]}

\newcommand{\kms}{\,km\,s$^{-1}$}

\definecolor{gold}{rgb}{0.85,.55,0}
\definecolor{orange}{rgb}{1,0.5,0}
\definecolor{cinza}{gray}{0.4}

\newcommand{\psqC}{PSQ\,J0330--0532}

\newcommand{\paren}[1]{\ensuremath{\left(#1\right)}} 		
\newcommand{\squares}[1]{\ensuremath{\left[#1\right]}}		

\title[2D stellar population and gas kinematics of the \psqC]{2D stellar population and gas kinematics of the inner kiloparsec 
of the post-starburst quasar SDSS\,J0330--0532}

\author[David Sanmartim, Thaisa Storchi-Bergmann and Michael S. Brotherton]{David Sanmartim$^{1}$\thanks{E-mail:
davidsanm@gmail.com}, Thaisa Storchi-Bergmann$^{1}$ and Michael S. Brotherton$^{2}$ \\
$^{1}$Universidade Federal do Rio Grande do Sul, IF, CP 15051, Porto Alegre 91501-970, RS, Brasil\\
$^{2}$University of Wyoming, Department of Physics and Astronomy, Laramie, WY, 82071, USA}
\begin{document}

\date{Accepted 2014 May 28. Received 2014 May 28; in original form 2013 October 12.}

\pagerange{\pageref{firstpage}--\pageref{lastpage}} \pubyear{2002}

\maketitle

\label{firstpage}

\begin{abstract}
We have used optical Integral Field Spectroscopy in order to map the star formation history of the inner kiloparsec of the Post-Starburst 
Quasar (PSQ) J0330--0532 and to map its gas and stellar kinematics as well as the gas excitation. PSQs are hypothesized to represent a stage 
in the evolution of galaxies in which the star formation has been recently quenched due to the feedback of the nuclear activity, as suggested 
by the presence of the post-starburst population at the nucleus. We have found that the old stellar population (age $\ge2.5$\,Gyr) dominates 
the flux at 5100\,\AA\ in the inner 0.26\,kpc, while both the post-starburst (100\,Myr\,$\le$ age $<$\,2.5\,Gyr) and starburst (age $< 100$\,Myr) 
components dominate the flux in a circumnuclear ring at $\approx$\,0.5\,kpc from the nucleus. 
With our spatially resolved study we do not have found any post-starburst stellar population in the inner 0.26\,kpc. On the other hand, we do see 
the signature of AGN feedback in this region, which does not reach the circumnuclear ring where the post-starburst population is observed. We thus 
do not support the quenching scenario for the \psqC. 
In addition, we have concluded that the strong signature of the post-starburst population in larger aperture spectra (e.g. from Sloan Digital Sky Survey)
is partially due to the combination of the young and old age components. 
Based on the M$_\rmn{BH}-\sigma_\rmn{star}$ relationship and the stellar kinematics we have estimated a mass for the supermassive black hole of $1.48 \pm 
0.66 \times$10$^7$ M$_\odot$.
\end{abstract}


\section{Introduction}
\label{sec:intro}

Post-Starburst Quasars (hereafter PSQs) are broad-lined active galactic nuclei (hereafter AGN) that show clear Balmer jumps and high-order Balmer absorption 
lines in their spectra, attributed to the contribution of A stars, characteristic of massive post-starburst stellar populations with age of a few hundred Myr. 
PSQs are hypothesized to represent a stage in the evolution of massive galaxies in which both star formation and nuclear activity have been triggered and are 
visible simultaneously before one or the other fades. \cite{brotherton99} report the particular case of PSQ UN\,J1025--0040 as the prototype for 
this class of objects, suggesting that it represents an intermediate stage in the evolution between an Ultra-Luminous Infrared Galaxy (ULIRG) and a Quasar.
The PSQs seem to represent a critical phase in the secular evolution of galaxies that links the growth of the stellar bulge and that of the SMBH. 

There are at least two possibilities to connect the presence of a post-starburst population to the nuclear activity: (1) a flow of gas towards the nucleus first 
triggers star formation in the circumnuclear region followed by an episode of nuclear activity which is triggered after hundreds of Myrs; in the meantime, the star 
formation may cease due to exhaustion of the gas \citep{sb01,davies07}; (2) the flow of gas towards the nucleus triggers star formation in the circumnuclear region 
and the nuclear activity, when triggered, quenches the star formation \citep{granato04,dimatteo05,hopkins06,canodiaz12}.

\cite{cales11} e \cite{cales13} recently performed photometric (with the HST/ACS camera) and spectroscopic studies (using integrated spectra obtained with the Keck/KPNO
telescope) for a sample of PSQs with $z\sim0.3$, which were selected spectroscopically from the SDSS. In these studies they show that the PSQ galaxies are an heterogeneous 
population, where both early-type and spiral galaxies can host AGN nuclei. Furthermore, they show that the PSQ galaxies present morphological characteristics and emission 
line ratios that suggest the presence of recent star formation together with a dominant post-starburst population. The near-infrared spectra \citep{wei13} also present 
characteristics of both ULIRGs as QSOs.
With the goal of investigating the nature of the connection between the post-starburst stellar population and nuclear activity in PSQs, we began a program to map 
the stellar population and the manifestations of nuclear activity in the inner few kpc of a sample of nearby PSQs using Integral Field Spectroscopy (IFS). 

In order to distinguish between the two scenarios above, we have used a sample of PSQs which have the clearest signatures of the post-starburst
population and redshifts lower than 0.1, since our goal is to resolve the stellar population on hundred of parsec scales.

In a previous exploratory study \citep[][hereafter Paper\,I]{sanmartim13} we have mapped the stellar population and the gas kinematics of the PSQ J0210-0903.  
We have used optical IFS obtained with the Gemini instrument GMOS using its Integral Field Unit (IFU), finding that old stars dominate the luminosity of the 
spectra at $\approx$\,4700\,\AA\ in the inner 0.3\,kpc (radius), but show also some contribution of post-starburst (intermediate age) population. Beyond this 
region (at $\approx$\,0.8\,kpc) the stellar population is dominated by both post-starburst and starburst population (young ionizing stars). The gas kinematics 
show a combination of rotation in the plane of the galaxy and outflows, observed with a maximum blueshift of $-670$\kms, resulting in a kinetic power for the 
outflow of $\dot{E}_\rmn{out} \approx 1.4-5.0 \times 10^{40}\,\rmn{erg}\,\rmn{s}^{-1} \approx 0.03-0.1\% \times L_\rmn{bol}$. 
This previous study has supported an evolutionary scenario in which the feeding of gas to the nuclear region has triggered a circumnuclear starburst 100's\,Myr ago, followed
by the triggering of the nuclear activity, producing the observed gas outflow which may have quenched further star formation in the inner 0.3\,kpc.

In this paper we present the results obtained from similar IFS observations of the PSQ SDSS J033013.26-053235.90 (hereafter \psqC), which is at a distance of 
approximately 50\,Mpc (from NED\footnote{The NASA/IPAC Extragalactic Database (NED) is operated by the Jet Propulsion Laboratory, California Institute of Technology, 
under contract with the National Aeronautics and Space Administration.}, for $H_\rmn{o} = 73.0 \pm 5$\,km\,sec$^{-1}$\,Mpc$^{-1}$), allowing the study of the 
spatial distribution of its stellar population and gas emission characteristics with a spatial resolution of 130\,pc. \psqC\ is one of the closest PSQs with $z<$\,0.1 
in our sample, and its spectrum clearly reveals the presence of high-order absorption lines of the series. The \psqC\ is hosted by a spiral galaxy with Hubble type 
Sb \citep{graham09} and its SDSS Petrosian absolute magnitude is $M_\rmn{i} = -22.22 \pm 0.05$. This target has more of a Seyfert-like luminosity rather than a quasar 
luminosity and its classification as a \qm{post-starburst quasar} is a generalized name for a broad-lined AGN with an intermediate-age stellar population. It has a 
companion galaxy distant $\sim$21\,kpc to the North and $\approx$16\,kpc to the West, probably on a collision course. \cite{greene06} and \cite{shen08}, have estimated 
the supermassive black hole masses of $\rmn{log}(\rmn{M}_\rmn{BH}) = 7.0 \pm 0.1$ and $\rmn{log}(\rmn{M}_\rmn{BH}) = 7.38 \pm 0.07$, respectively, in solar units. 

This paper is organized as follows. In Section \ref{sec:obs_reduc} we describe the observations and reduction processes. In Section \ref{sec:results} 
we report our results on the stellar population (Sec. \ref{sec:population}), stellar kinematics (Sec. \ref{sec:stellarkinematics}), emitting gas flux 
distribution and excitation (Sec. \ref{sec:emitting}), gas kinematics (Sec. \ref{sec:kinematics} and channel maps for the emitting gas \ref{sec:slices}).  
In Section \ref{sec:discussions} we discuss and interpret our results. In sections \ref{sec:mass} and \ref{sec:outflowrate} we present an estimate to 
the mass of the narrow line-emitting gas and the mass outflow rate, respectively. In Section \ref{sec:conclusions} we present a summary of our results as well 
as our conclusions.

\begin{figure*}
\begin{minipage}[c]{1.0\textwidth}
\begin{center} 
  \includegraphics[scale=0.60,angle=0]{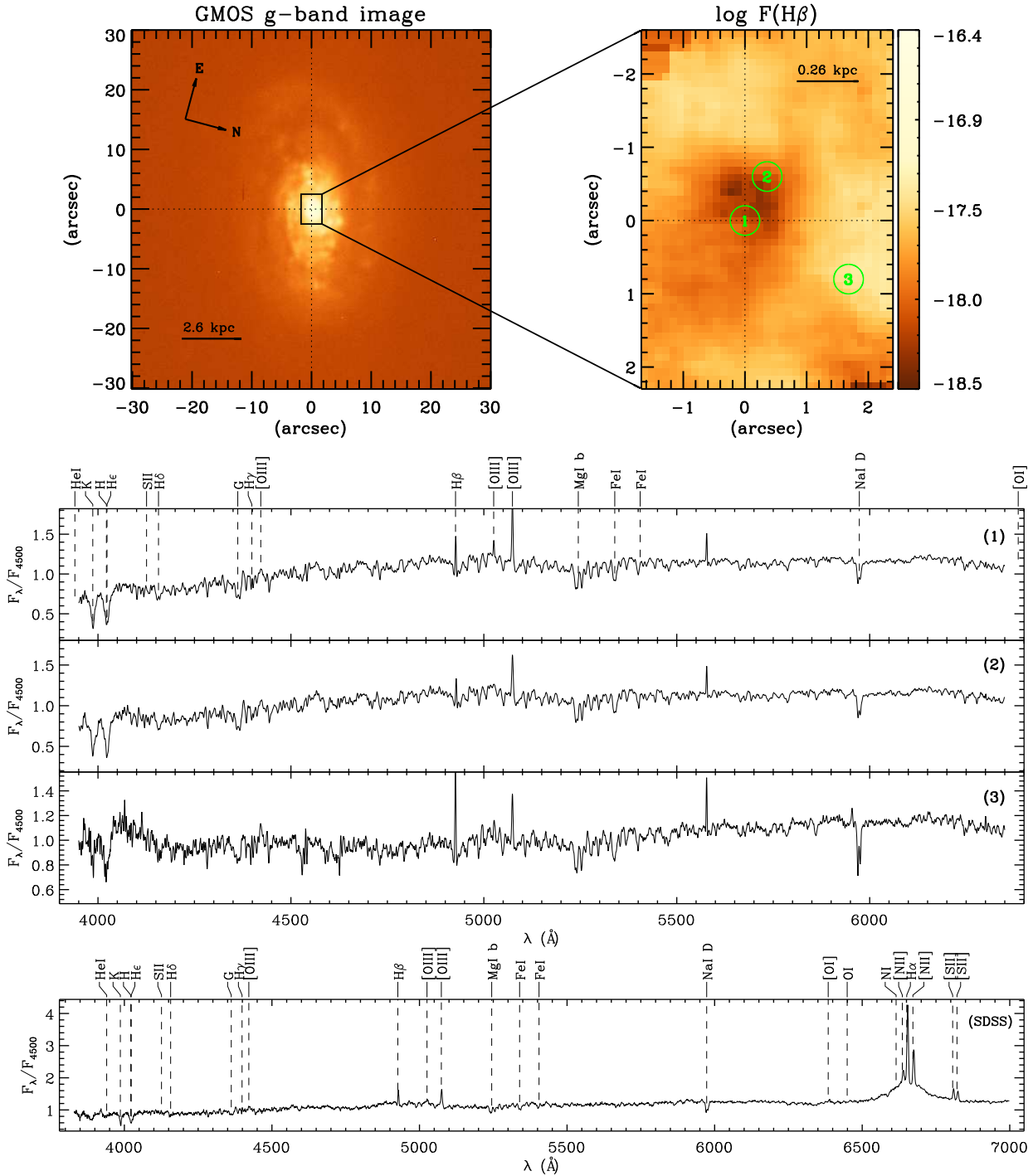}
  \caption{
           The top left panel shows the GMOS g-band acquisition image of the \psqC\ and a box that represents the field-of-view of the GMOS-IFU observations. The top right
           panel shows the flux map in H$\beta$ with 3 regions from which the spectra shown bellow were integrated. Regions 1, 2 and 3 have an aperture diameter of 
           0\farcs4 ($\sim$100\,pc). The SDSS spectrum of the \psqC\ is shown in the bottom panel with the same wavelength range of our IFU spectra.
           }
  \label{fig:galspec}
\end{center}
\end{minipage}
\end{figure*}

\section{Observations and data reduction}
\label{sec:obs_reduc}

Two-dimensional spectroscopic data of the \psqC\ were obtained on December 14 and 15, 2010, using the Integral Field Unit of the Gemini Multi-Object Spectrograph 
(GMOS-IFU) at the Gemini South telescope under the Gemini project GS-2010B-Q-24. The observations were obtained in one-slit mode, covering a field of view (FOV) 
of 3.0\,$\times$\,5.0 arcsec$^2$. We have used the B600$\_$G5307 grating, which results in a spectral range of $\sim$ $3950-6350$\,\AA\ in the rest frame and a 
wavelength sampling of $\sim$0.5\,\AA\ pixel$^{-1}$ at a instrumental resolution of 1.6\,\AA\ at 5200\,\AA, corresponding to $\sigma_\rmn{\textsc{inst}}$ of 
$\approx40$\kms. The total exposure time was 22000\,s, taken from 8 individual exposures of 2750\,s each. The mean seeing of the observations was in the range 
of 0\farcs5--0\farcs6 during the first night and of 0\farcs3--0\farcs5 during the last night, corresponding to an average spatial resolution of $\sim$\,0\farcs5. 
This corresponds to a spatial resolution of $\sim$130\,pc at the distance of the galaxy.  

Data reduction was accomplished using generic \textsc{iraf}\footnote{\textsc{iraf} is distributed by the National Optical Astronomy Observatories, which is 
operated by the Association of Universities for Research in Astronomy, Inc. (AURA) under cooperative agreement with the National Science Foundation.} tasks 
and specific ones developed for GMOS data in the \textsc{gemini.gmos} package. The basic reduction steps are similar to the ones detailed in Paper\,I.
The final datacube contains 1750 spectra each corresponding to an angular coverage of 0\farcs1\,$\times$\,0\farcs1 or 26\,$\times$\,26\,pc$^2$ at the distance of 
the galaxy. Cosmic rays were cleaned from the data before sky-subtraction with Laplacian cosmic ray identification routine \textsc{lacosmic} \citep{vandokkum01}. 
The spectra were corrected for reddening due to the interstellar Galactic medium using the \textsc{iraf} routine \textsc{noao.onedspec.deredden} for the $V$-band 
extinction $A_\rmn{V}=0.155$; its value was calculated using the NED extinction calculator, which uses the \cite{schlegel98} Galactic reddening maps.

\begin{figure*}
\begin{minipage}[c]{1.0\textwidth}
\centering 
\includegraphics[scale=0.56,angle=0]{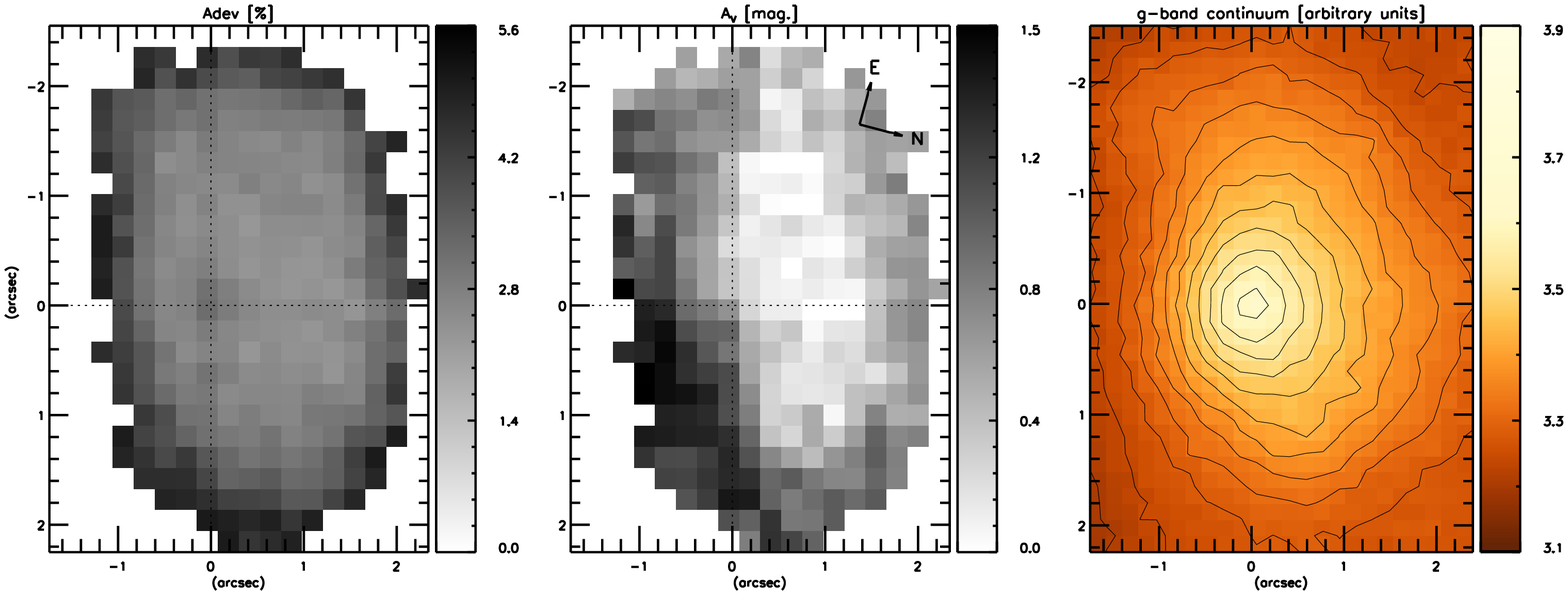}
\caption{This figure shows the map of the $adev$\,(left), the map of the intrinsic extinction in the V passband $A_\rmn{V}$ (middle) and the map of the continuum 
         image in the g-band, with overploted contours (right).}
 \label{fig:reddening}
\end{minipage}
\end{figure*}

In the top-left panel of Figure \ref{fig:galspec} we present the Gemini-GMOS $g$-band acquisition image of the PSQ, where the box represents the FOV of the IFU 
observations. In the top-right panel we present the H$\beta$ flux distribution -- after subtraction from the continuum contribution -- within the IFU 
FOV. The set of 3 circular regions, illustrated by circles in the figure, have radii of 0\farcs2, from which we have extracted representative integrated spectra shown 
in the panels bellow. The region 1 is centered at the nucleus (defined to be the location of the peak of the continuum), while regions 2 and 4 are centered at 0\farcs7
and 1\farcs9 ($\sim$190 and 490\,pc) from the nucleus. It can be seen that the narrow H$\beta$\ emission increases outside the nucleus. In the bottom panel we show the SDSS 
spectrum of the galaxy for an wider spectral range, corresponding to an aperture of 3 arcsec.

\section{Results}
\label{sec:results}

\subsection{Stellar population}
\label{sec:population}

In order to study the stellar population distribution, we performed spectral synthesis using the {\small STARLIGHT} code \citep{cid04,cid05,cid09}, which searches for 
the linear combination of $N_*$ simple stellar populations (SSP), from a user-defined base, that best matches the observed spectrum. Basically, the code fits an observed 
spectrum $O_\lambda$ solving the following equation for a model spectrum $M_\lambda$ \citep{cid04}:

\begin{equation}
M_\lambda={M_{\lambda}}_{0} \left( {\sum_{j=1}^{N_*} {x_j}{b_{j,\lambda}}} \right) {r_\lambda} \otimes G(\upsilon_*,\sigma_*), 
\label{eq:M_lambda}
\end{equation}
where
${M_{\lambda}}_{0}$							is the synthetic flux at the normalization wavelength,
$\textbf{x}$								is the population vector, whose components represent the fractional contribution
									of the each SSP to the total synthetic flux at $\lambda_0$, 
$b_{j,\lambda} \equiv 
{L_{\lambda}^{SSP}(t_j,Z_j)}/{L_{\lambda_0}^{{SSP}}(t_j,Z_j)}$	is the spectrum of the $j$th SSP, with age $t_j$ and metallicity $Z_j$, normalized at $\lambda_{0}$,
$r_\lambda \equiv 10^{-0.4(A_\lambda-A_{\lambda_{0}})}$			is the reddening term, and 
$G(\upsilon_*,\sigma_*)$ 						is the Gaussian distribution, centred at velocity $\upsilon_*$ with dispersion $\sigma_*$, used to model 
									the line-of-sight stellar motions.
The reddening term is modeled by {\small STARLIGHT} as due to foreground dust and parametrized by the V-band extinction $A_\rmn{V}$ so that all components are equally reddened
and to which we have adopted the Galactic extinction law of \cite{cardelli89} with $R_\rmn{V}=3.1$.

The match between model and observed spectra is carried out minimizing the following equation \citep{cid04}:
\begin{equation}
\chi^2= {\sum_{\lambda} \left[ (O_\lambda - M_\lambda) w_\lambda \right]^2}, 
\label{eq:qui2}
\end{equation}
where $w_\lambda$ is the weight spectrum, defined as the inverse of the noise in $O_\lambda$, whose emission lines and spurious features are masked out by fixing $w_{\lambda}=0$ 
at the corresponding $\lambda$. The minimum to equation \ref{eq:qui2} corresponds to the best parameters of the model and the search for them is carried out with a simulated 
annealing plus the Metropolis scheme. A detailed discussion of the Metropolis scheme applied to the stellar population synthesis can be found in \cite{cid01}.

We constructed the spectral base with the high spectral resolution evolutionary synthesis models by \cite{bc03} (BC03), where the SSPs cover 11 ages, $t=1.0\times10^6$, 
$5.0\times10^6$, $2.5\times10^7$, $1.0\times10^8$, $2.9\times10^8$, $6.4\times10^8$, $9.1\times10^8$, $1.4\times10^9$, $2.5\times10^9$, $5\times10^9$ and $1.1\times10^{10}$
\,yrs, assuming solar ($Z_\odot=0.02$) and super-solar metallicity ($Z=2.5\,Z_\odot$). We have used the SSP spectra constructed from the STELIB library \citep{leborgne03}, 
Padova-1994 evolutionary tracks and \cite{chabrier03} Initial Mass Function (IMF). In order to account for the AGN featureless continuum (FC) a non-stellar component was 
also included, represented by a power law function (fixed as $F_\nu \propto \nu^{-1.5} $). In accordance to \cite{cid04} we have binned the contribution of the population in a reduced 
population vector with only three stellar population components (SPCs) corresponding to three age ranges: \textsc{young} ($x_{\rm Y}: t < 100$\,Myrs), \textsc{intermediate} 
($x_{\rm I}$: $100\,\rmn{Myr} \leq t < 2.5\,\rmn{Gyr}$) and \textsc{old} ($x_{\rm O}$: $t \geq 2.5$\,Gyr). 

\begin{figure*}
\begin{minipage}[c]{1.0\textwidth}
\centering 
\includegraphics[scale=0.5,angle=0]{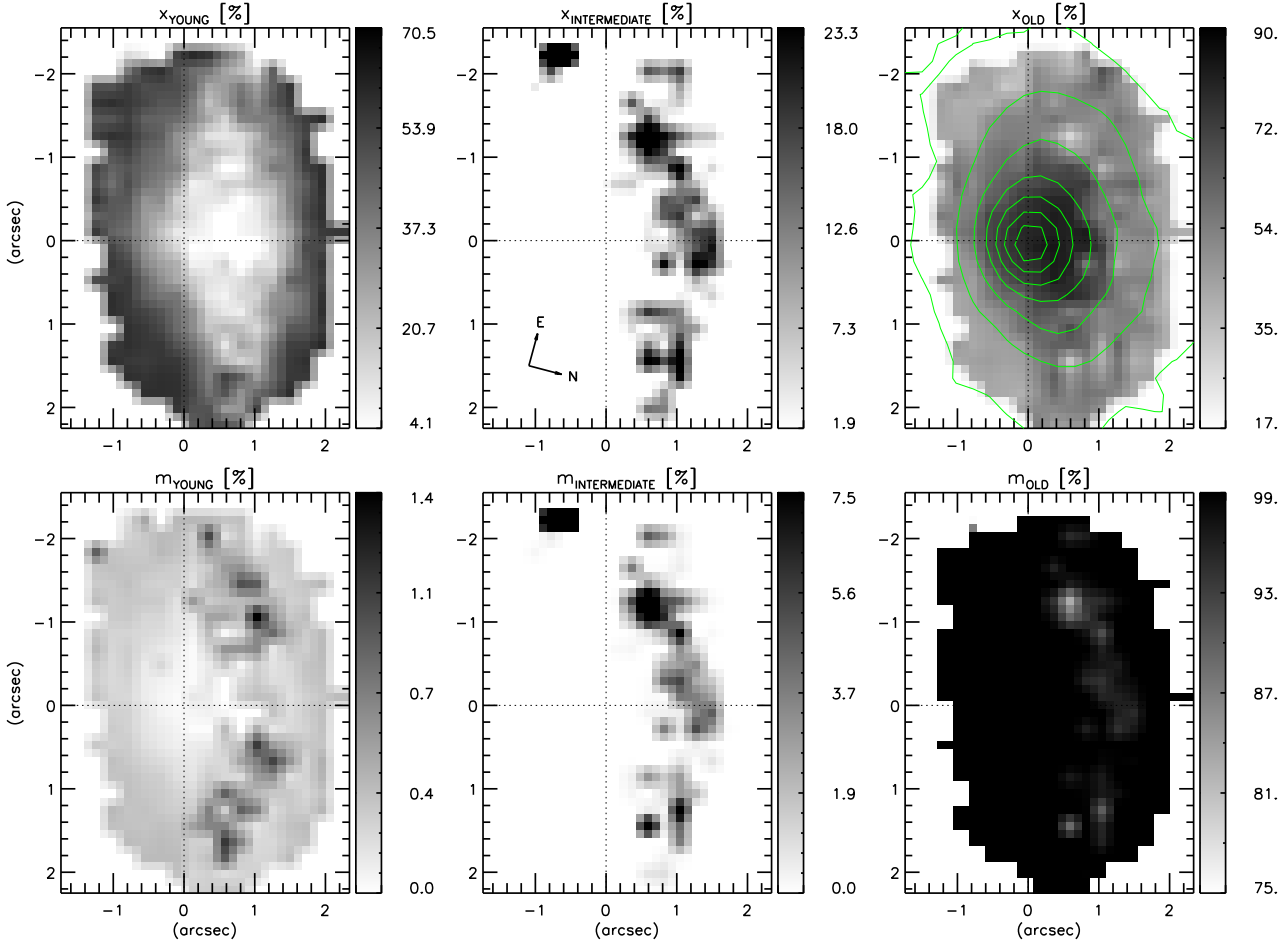}
\caption{Spatial distributions of the per cent contribution of the stellar population components. At the top panels we show the maps of the population vector in per cent 
         units of the flux at 5100\,\AA\ for ages \textsc{young} ($x_\rmn{\textsc{young}}$), \textsc{intermediate} ($x_\rmn{\textsc{intermediate}}$) and \textsc{old} 
         ($x_\rmn{\textsc{old}}$). At the bottom panels we show the percent mass fraction for each age. The contours in the $x_\rmn{\textsc{old}}$ map are from the 
          continuum image.}
 \label{fig:synthesis}
\end{minipage}
\end{figure*}

\begin{figure*}
\begin{flushleft}
\begin{minipage}[c]{1.0\textwidth}
  \includegraphics[scale=0.345,angle=0]{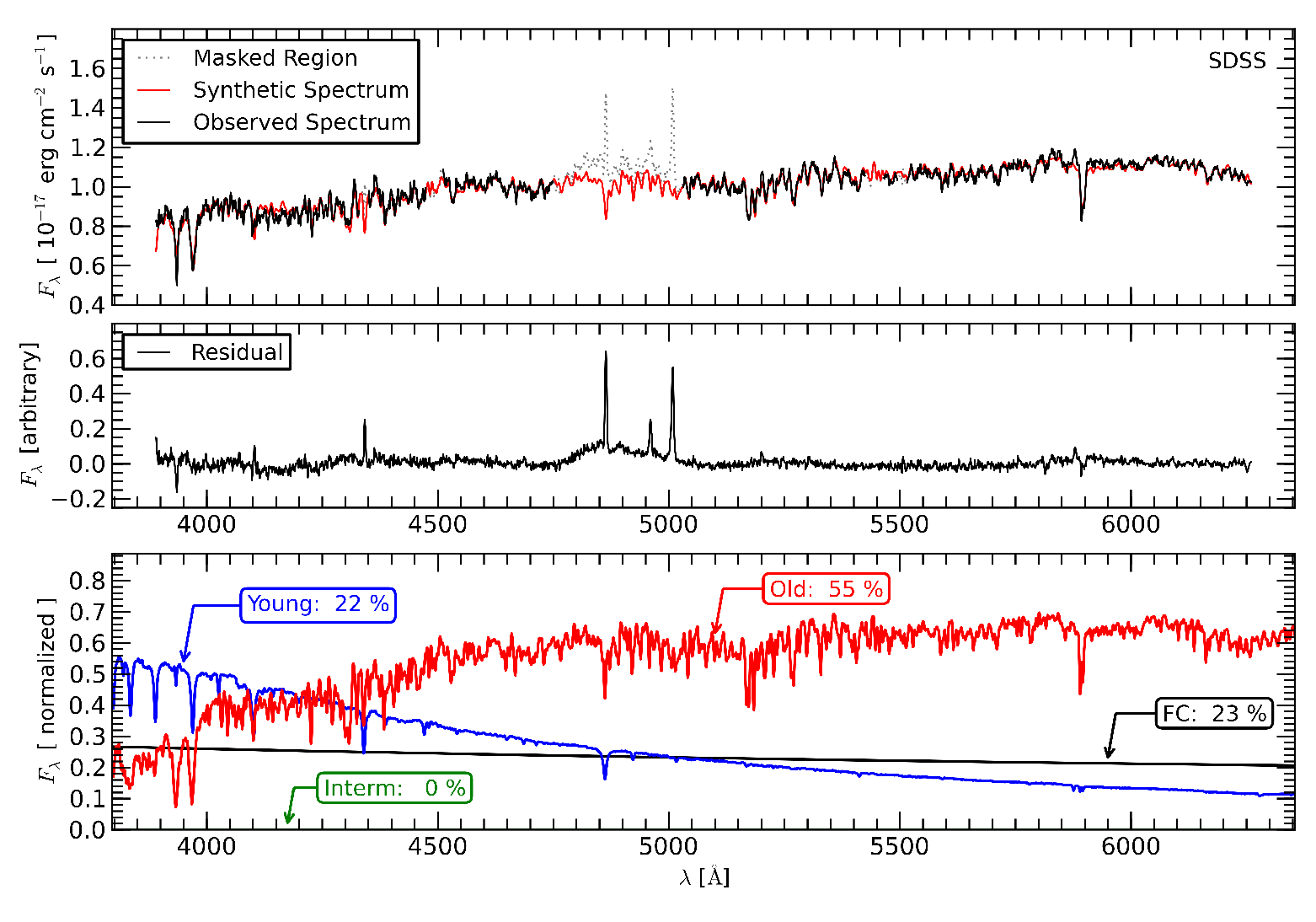}
  \includegraphics[scale=0.345,angle=0]{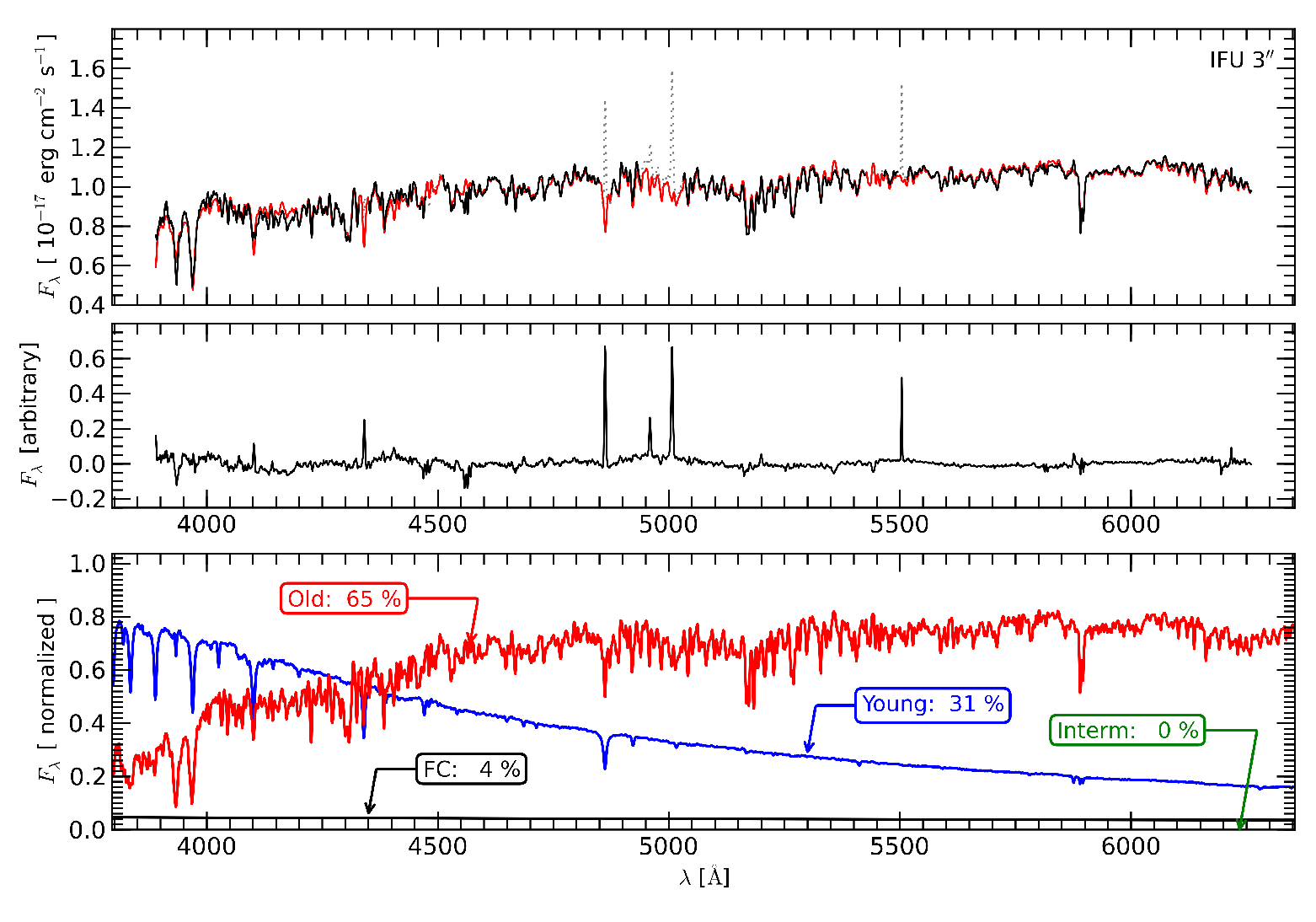}
    \caption{Synthesis results for the SDSS spectrum (left) and for the IFU spectrum (right) integrated within an aperture radius of 1\farcs5 around the nucleus. The black 
             solid line at the top of each panel shows the observed spectrum. The red solid line shows the fitted model, while the dotted line shows the masked regions of 
             the spectrum (emission lines or spurious features). At the middle panels we show the residual spectrum, obtained from the difference between the observed and 
             modelled spectra. At the bottom panels we show the spectra corresponding to each age bin (\textsc{young} -- blue; \textsc{intermediate age} -- green; \textsc{old} 
             -- red) and the FC component \textsc{FC} -- black), both scaled to its percent contribution to the total continuum light at 5100\,\AA.}
\label{fig:sdss+ifu}
\end{minipage}
\end{flushleft}
\end{figure*}

The goodness of the fit is measured by the $adev$ parameter, which gives the per cent mean difference between the modelled and observed spectrum. In Figure 
\ref{fig:reddening}, we present the $adev$ parameter (left) and the intrinsic extinction in the V passband $A_\rmn{V}$ (right). The values of the $adev$ are $\leq5$ per cent, 
which we consider a limiting value for good fits. We have used this value to exclude from our stellar population maps (Figure \ref{fig:synthesis}) the pixels for 
which the corresponding $adev$ is larger than 5 per cent. The extinction reaches the highest values, up to $A_\rmn{V}$=1.5, in an elongated half ring along the 
second and third quadrants of the mapped field.  

In Figure \ref{fig:synthesis} we show the percent flux contribution at 5100\,\AA\ of each SPC (top panels) and the percent mass fraction (bottom panels). Within 
the inner $\approx$1\,arcsec (0.26\,kpc) the contribution of the old SPC dominates the continuum, representing up to$\approx$85 per cent of the flux at 5100\,\AA. The 
peak contribution is displaced $\approx$0\farcs4 to the North of the nucleus. The contribution of the old SPC decreases to $\approx$35 per cent towards the borders 
of the field, where the young SPC dominates, contributing with $\approx$65 per cent of the flux at 5100\,\AA. In an elongated half ring at the first and fourth
quadrants (of the top center panel) the intermediate age SPC contributes with up to $\approx$23 per cent of the continuum at 5100\,\AA. When we consider the 
mass fraction, the old SPC dominates everywhere with at least $\approx$90 per cent of the mass. The intermediate age SPC presents a significant contribution to the
mass, reaching up to $\approx$7 per cent in the half ring, while the young SPCs contribute with no more than $\approx$1.5 per cent of the total stellar mass.

In Figure \ref{fig:sdss+ifu} we show the results of the synthesis for the SDSS spectrum and for an IFU integrated spectrum within the SDSS aperture radius (1\farcs5). 
The black solid line at the top of each panel shows the observed spectrum. The red solid line shows the fitted model, while the dotted line shows the masked regions 
of the spectrum (emission lines or spurious features). In the middle panels we show the residual spectrum, obtained from the difference between the observed and modelled 
spectrum. In the bottom panels we show each SPC spectrum weighted by its fractional contribution to the flux at 5100\,\AA (\textsc{young} -- blue; \textsc{intermediate 
age} -- green; \textsc{old} -- red) and the FC component \textsc{FC} -- black). The best fits of SDSS and IFU integrated spectra both present 
no contribution of the intermediate age SPCs, although the intermediate age population signatures seem to be present in the observed spectra. The old SPC contributes with $\approx$55 
(SDSS) and $\approx$65 (IFU) per cent, while the FC component plus young SPC contributes with $\approx$45 and 35 per cent, respectively. The stellar population 
synthesis for both the SDSS and integrated IFU spectra thus approximately agree, presenting a difference of $\approx$10 per cent in the contribution of the old SPC and FC plus young 
SPCs. The $adev$ parameter for both modelled spectra are $\approx$1.8 per cent, which corresponds to a very good fit. Although the stellar population synthesis method 
does not provide error estimates for the SPCs contributions, we can use this difference ($\approx$10 per cent) as an error estimates for the SPC contribution. 
We noted a difference between the SDSS and the integrated IFU spectrum: a broad H$\beta$ component which is present only in the SDSS spectrum. We attribute 
this difference to a possible variation of the the broad line, which seems to have practically disappeared in our spectra, obtained at a later date than that of 
the SDSS spectrum. Such variation is typical of the Broad Line Region and has been reported in many previous studies, The \qm{flat-top} profile can be a signature
that this broad line may originate in the outer parts of as accretion disk \citep[e.g.,][]{thaisa97,Ho00,strateva06,strateva08}. 
 
\subsection{Stellar Kinematics}
\label{sec:stellarkinematics}

\begin{figure*}
\begin{minipage}[c]{1.0\textwidth}
\centering 
\includegraphics[scale=0.42, angle=0]{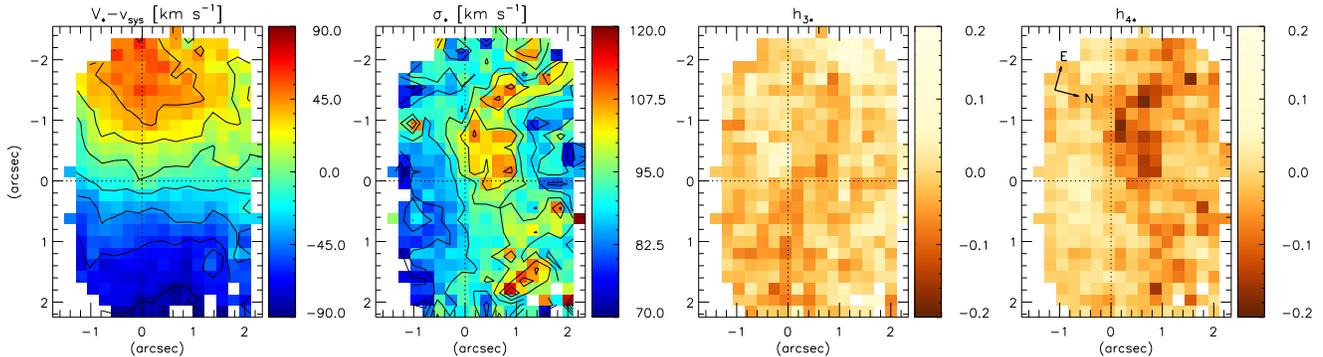}
 \caption{Two-dimensional maps of the stellar kinematics obtained from the pPXF method. In the two leftmost panels we show the centroid velocity and the dispersion velocity
          maps. In the two rightmost panels, the Gauss-Hermite moments $h_3$ and $h_4$ are shown. The mean uncertainties are 9\kms\ for the centroid velocities, 7\kms\
          for the dispersion velocities and 0.02 for the Gauss-Hermite moments. The vertical black dotted line shows the orientation of the line of the nodes, obtained 
          from the rotation model.
         }
 \label{fig:starkinematics}
\end{minipage}
\end{figure*}

We obtained the stellar line-of-sight velocity distributions (LOSVD) by fitting the absorption features of the spectra, which include, in particular, the triplet 
of magnesium MgI\,b $\lambda\lambda$5167, 5173, 5184\,\AA\ and the absorption doublet of sodium NaI\,D $\lambda\lambda5890,\,5896$\,\AA\ using the penalized Pixel-Fitting
\footnote{http://www-astro.physics.ox.ac.uk/~mxc/idl, last accessed October 28, 2012.} (pPXF) method of \cite{cappellari04}. In summary, the pPXF recovers 
parametrically the best fit to an observed spectrum by convolving stellar template spectra with a defined LOSVD ($\mathcal{L}(v)$), which is represented by 
a Gauss-Hermite (GH) series that can be written as \citep{marel93}:
\begin{equation}
\mathcal{L}(v) = \frac{e^{-(1/2)y^2}}{\sigma\sqrt{2\upi}} \squares{\, 1 + {\sum^{M}_{m=3}}\,h_m H_m(y) \,} \label{eq:LOSVD}\,
\end{equation}
where $y=(v-V_*)/\sigma_*$, $v=c\ln\lambda$, $c$ is the speed of light, $V_*$ is the stellar centroid velocity, $\sigma_*$ is the stellar velocity dispersion, 
$H_m$ are the Hermite polynomials and $h_m$ are the GH moments \citep{cappellari04}. The pPXF routine, constructed in IDL, outputs the stellar kinematics ($V_*$ and $\sigma_*$) 
and the high-order Gauss-Hermite moments $h_3$ and $h_4$. We have used a library of stellar templates constructed from the SSP models of \cite{bc03}.

In Figure \ref{fig:starkinematics} we present the two-dimensional maps of the stellar kinematics resulting from the application of the pPXF routine. The nucleus is 
identified by the intersection of the perpendicular dotted lines. In the two leftmost panels we show the centroid velocity and velocity dispersion fields of the 
stellar kinematics. The systemic velocity $V_\rmn{sys}=3965\pm5$\kms\ was obtained from the rotation model employed to reproduce the velocity field (see Sec 
\ref{sec:rotation}) and was subtracted from the centroid velocity field. The vertical black dotted line shows the approximated orientation of the line of the nodes, 
obtained from the same fit.

The velocity field shows a clear rotation pattern, with redshifts to the East and blueshifts to the West. The amplitude of the rotation is $\approx$70\kms and its 
kinematic center is located at the same position of the continuum peak, given the uncertainties and spatial resolution. The mean uncertainties in the centroid 
velocities are of $\approx$9\kms. The velocity dispersion values are in the range 70 to 110\kms. A region of low velocity dispersion ($\sigma_*\approx80-90$\kms) 
encircles a nuclear region of higher $\sigma_*$ values ($\approx$110\kms), which is displaced from the nucleus 0\farcs5 to the north-east. The stellar velocity dispersion 
has a mean uncertainty of $\approx$7\kms within the inner 1\,arcsec (radius), increasing up to 15\kms\ at the borders of the field. Pixels for which the SNR decreases 
to below 25 were not considered in the fit, since a value $\gtrsim 25$ is required to obtain reliable velocity measurements.
In the rightmost panels of Figure \ref{fig:starkinematics} we show the $h_3$ and $h_4$ moments, respectively. The Gauss-Hermite moments measure the deviations 
of the line profile from a Gaussian. While the $h_3$ quantify the asymmetric deviations, the $h_4$ measures the symmetric deviations. Most $h_3$ and $h_4$ values 
are in the range $-0.04$ to $+0.04$. The lowest values in the $h_3$ moments reach $\approx-0.1$, while the highest are of $\approx+0.1$, presenting an irregular 
distribution in the FOV. The highest values of $h_4$ ($\approx0.1$), on the other hand, seem to be distributed in an elongated half ring to the South-Southeast
surrounding the lowest values ($\approx-0.15$) of the $h_4$ at the nucleus and to the Northeast. This distribution of the $h_4$ moments is correlated with the 
distribution of the velocity dispersion values: the region with the lowest values of $\sigma_*$ has the highest values of the $h_4$ moments, while the region with
the highest values of $\sigma_*$ has the lowest values of $h_4$. 

\subsection{Emitting Gas}
\label{sec:emitting}

\subsubsection{Emission line fits and error estimates}
\label{sec:measure+error}

The kinematics and flux distributions of the emitting gas were obtained by fitting Gaussian curves to the line profiles of \oiii\footnote{We denote the 
\oiii$\lambda5007$ emission line simply by \oiii.} and H$\beta$. These profiles were measured using an IDL routine that solves for the best solution of parameters 
of the Gaussian function using the nonlinear least-squares Levenberg-Markwardt method. The IDL routine used is similar to that described in Paper\,I. We have 
performed the fit to the spectrum of each pixel of the datacube in order to obtain the spatial distribution of the emission-line fluxes, velocity dispersions 
and centroid velocities. 

Error estimates were obtained using Monte Carlo simulations adding to the original spectrum an artificial Gaussian noise normally distributed, whose mean 
is equal to zero and standard deviation equal to one. We have performed the simulations with 1000 realizations, computing the best-fitting parameters of 
each realization so that we have the standard deviation of the best model in the end of the simulation. In this way, we have obtained the error estimates 
for the free parameters of the Gaussian function: the centroid velocity, velocity dispersion and flux of each emission line. The error values in each measured 
parameter are given together with the description of the corresponding maps in the following sections. The errors for the emission line ratio \oiii/H$\beta$ 
were derived from the errors in the \oiii\ and H$\beta$ emission lines. 

\begin{figure*}
\begin{minipage}[c]{1.0\textwidth}
\centering 
\includegraphics[scale=0.53,angle=0]{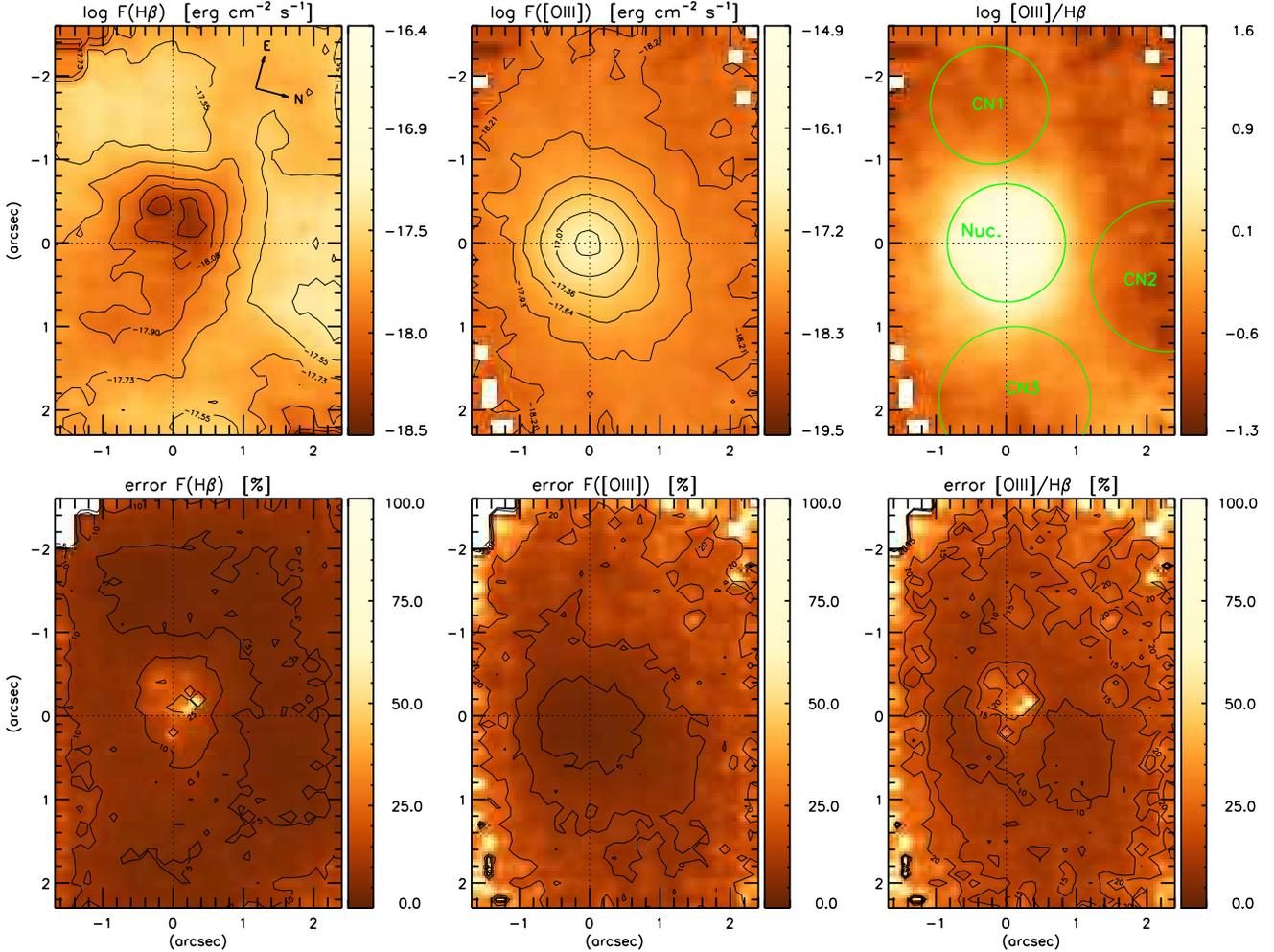}
     \caption{Integrated fluxes for H$\beta$ (top left) and \oiii\ emission lines (top central panel), given in logarithmic units of erg\,cm$^{-2}$\,s$^{-1}$. Top right 
              panel: emission-line ratio \oiii/H$\beta$ in logarithmic scale. The labels Nuc., CN1, CN2 and CN3 identify the regions from where the line ratios shown in 
              the BPT diagrams were extracted (see Sec. \ref{sec:dis.excitation}). In the bottom panels the corresponding per cent errors, obtained using Monte Carlo 
              simulations, are shown. For both emission lines as well as for the emission line ratios the mean uncertainties are less than 10 per cent for the most part 
              of the mapped field.}
 \label{fig:flux}
\end{minipage}
\end{figure*}

\subsubsection{Emission-line flux distributions}
\label{sec:flux}

In Figure \ref{fig:flux} we present the flux distributions in H$\beta$ (top left panel) and \oiii$\lambda$5007 (top central panel) emission lines, both given in 
logarithmic scale and erg\,cm$^{-2}$\,s$^{-1}$ units. In the top right panel we show the spatial distribution of the emission line ratio \oiii/H$\beta$ in logarithmic 
scale. The location of the maximum brightness of the \oiii\ flux distribution -- and of the highest emission line ratio \oiii/H$\beta$ -- coincides with the position 
of the nucleus. This position was determined as the centroid of the flux distribution in the continuum, obtained by collapsing the datacube between two continuum 
wavelengths. We have adopted this position as the galaxy nucleus and its location in the figures is identified by the intersection of perpendicular dotted lines. 

The H$\beta$ map shows a flux distribution in which the highest values are distributed in a partial ring surrounding the nucleus from East through North to West.
Assuming that the ring is in the plane of the galaxy, we estimate its radius to be $\approx$500\,pc. 
To the Southwest of the nucleus, the ring seems to disappear, at a location where the stellar population synthesis has shown the highest values of $A_\rmn{V}$.
At about 0\farcs5 to the East and North of the nucleus there are two compact regions where the H$\beta$ emission is very weak. The errors in the H$\beta$ emission 
line fluxes are less than $\approx$\,6 per cent in the partial ring. Inside the ring, the maximum error in the H$\beta$ flux is $\approx$25 per cent. Only in a few 
pixels close to the nucleus the errors reach up to $\approx$75 per cent. On average, in the other regions of the mapped field the errors in H$\beta$ are less than 
$\approx$12 per cent.

The flux distribution of the \oiii\ emission line (top central panel of Fig. \ref{fig:flux}) is more symmetrically distributed around the nucleus, with the flux
values decreasing smoothly and almost radially towards the borders of the mapped field. The errors in the \oiii\ fluxes increase with distance from the nucleus. 
Within the inner 0\farcs5 the errors are less than $\approx$5 per cent, while between 0\farcs5 and 1\farcs0 the errors are smaller than $\approx$10 per cent. 
Beyond this regions, up to the borders, the errors increase to $\approx$10--20 per cent. Only at the  borders of the mapped field, the errors are between 
$\approx$40 and $\approx$75 per cent.

\subsection{Gas Kinematics}
\label{sec:kinematics}

\begin{figure*}
\begin{minipage}[c]{1.0\textwidth}
\centering 		
\includegraphics[scale=0.43, angle=0]{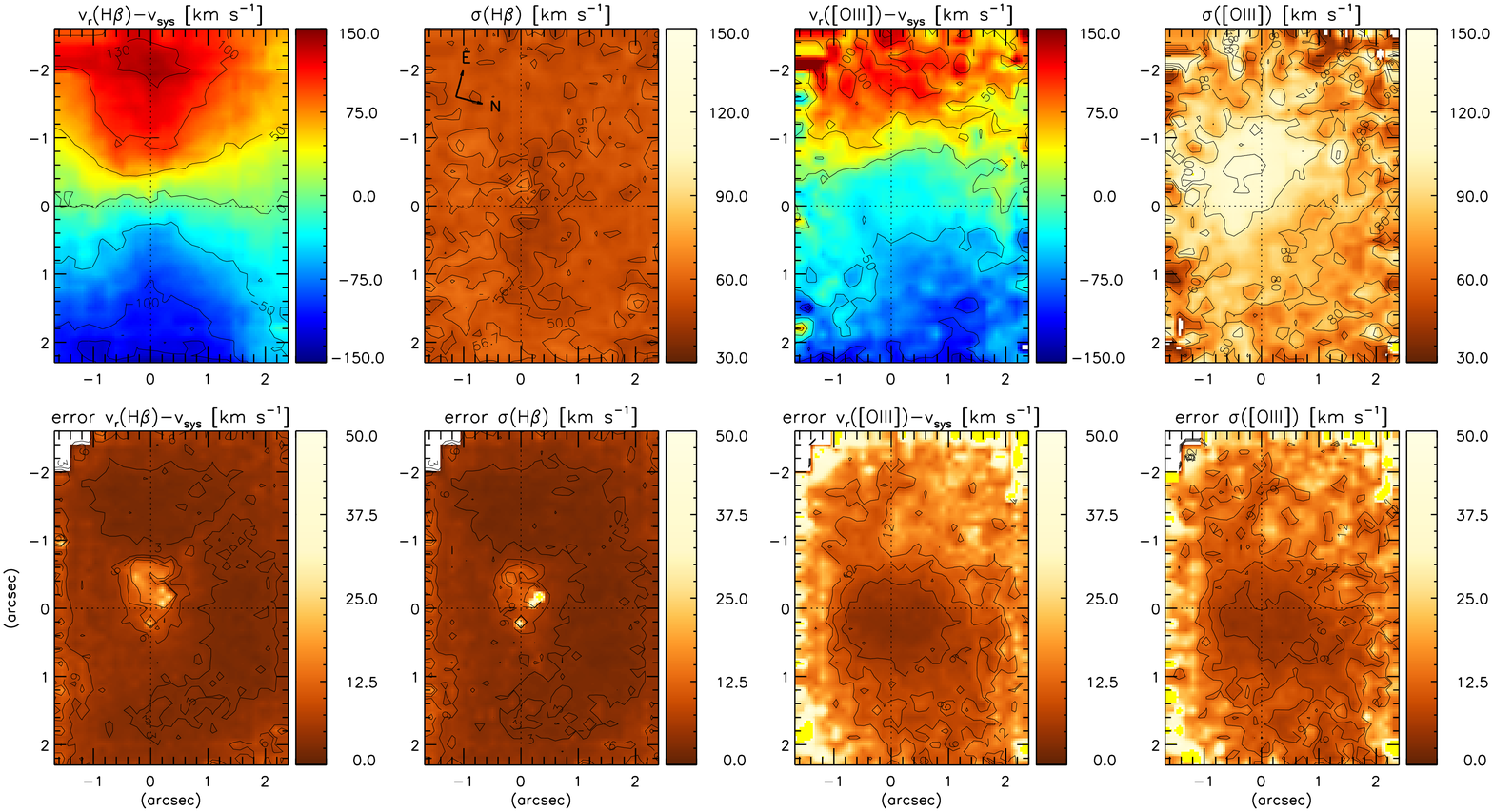}
 \caption{
          Left: Centroid velocity maps of \oiii\ and H$\beta$ in units of km\,s$^{-1}$. Right: Velocity dispersion maps. Both have been obtained from the fit of GH 
          series to the emission-line profiles. The mean uncertainties are between $\approx$6\,km\,s$^{-1}$ and 15\,km\,s$^{-1}$ for both the velocity 
         fields and velocity dispersion maps for \oiii. For H$\beta$ the mean uncertainties are between $\approx$10\,km\,s$^{-1}$ and 25\,km\,s$^{-1}$.
         }
 \label{fig:cinematica}
\end{minipage}
\end{figure*}

We have obtained the gas kinematics by using a Gaussian function to fit the H$\beta$ and \oiii\  emission lines (Sec. \ref{sec:flux}). Figure \ref{fig:cinematica} 
shows the velocity fields obtained from these fits. In the top panels we present the centroid velocity and velocity dispersion maps, while in the bottom panels 
we show the corresponding error maps. Centroid velocities are shown relative to the systemic velocity of the galaxy (3965$\pm5$\,km\,s$^{-1}$).

The H$\beta$ velocity field (top-left panel) presents a rotation pattern with the line of the nodes oriented  approximately along East-West, with positive values to 
the East and negative values to the West, and a velocity amplitude of $\approx$130\,\kms. The errors in the velocity values are smaller than 6 per cent over most 
the FOV. At the two compact regions of weak H$\beta$ emission at $\approx$ 0\farcs5 from the nucleus (see Fig.\,\ref{fig:flux}), the errors in the centroid velocity 
increase to $\approx$20 per cent. In the half ring of high H$\beta$ emission, the errors are lower than 3 per cent, while at the borders of the FOV the errors are 
of $\approx$12 per cent. In the other parts of the FOV the errors are on average less than 6 per cent.

The H$\beta$ velocity dispersion map (second top panel of Fig.\,\ref{fig:cinematica}) shows values in the range 40--60\,\kms, indicating that the random non-circular 
motions are less important than the rotation motion for the gas. The lowest velocity dispersion values ($\approx$40\,\kms) are observed near the nucleus, while the 
highest values ($\approx$60\,\kms) are farther away, in the ring of enhanced H$\beta$ emission. The error values and their distribution are similar to those of the 
centroid velocities.

The \oiii\ centroid velocity field (third top panel of Fig.\,\ref{fig:cinematica}) also shows a pattern that suggests rotation, but it is disturbed by non circular 
motions. The \oiii\ kinematics seems to be a mixture of rotation and outflow evidenced by the excess blueshifts at and around the nucleus when compared with the 
H$\beta$ velocity field. But the \oiii\ rotation amplitude and orientation of the line of nodes are similar to  that of H$\beta$. The average errors in the \oiii\ 
centroid velocities are lower than 17 per cent for most regions. Close to the limits of the mapped field the errors increase to approximately 30 per cent.

The \oiii\ velocity dispersion values (top-right panel of Fig.\,\ref{fig:cinematica}) range between 80 and 130\,\kms, with the highest values (which are comparable 
to the gas rotation amplitude) being observed at a region $\approx$0\farcs5 Southeast of the nucleus and extending by $\approx$1 arcsec in all directions. 
The error values and their distribution are similar to those observed for the centroid velocity map.

\subsection{Channel maps}
\label{sec:slices}

In order to map the velocity field of the emitting gas throughout the emission line profile -- and not only at the central wavelength, as done in the centroid 
velocity maps, we have obtained channel maps by integrating the flux distribution in different slices of velocity along the \oiii\ and H$\beta$ emission-line 
profiles. We have integrated the flux in each velocity channel after subtraction of the continuum contribution evaluated as the average between the fluxes in 
the continuum at both sides of the profile.   

\begin{figure*}
\begin{minipage}[t]{1.0\textwidth}
\centering 
  \includegraphics[scale=0.96,angle=0]{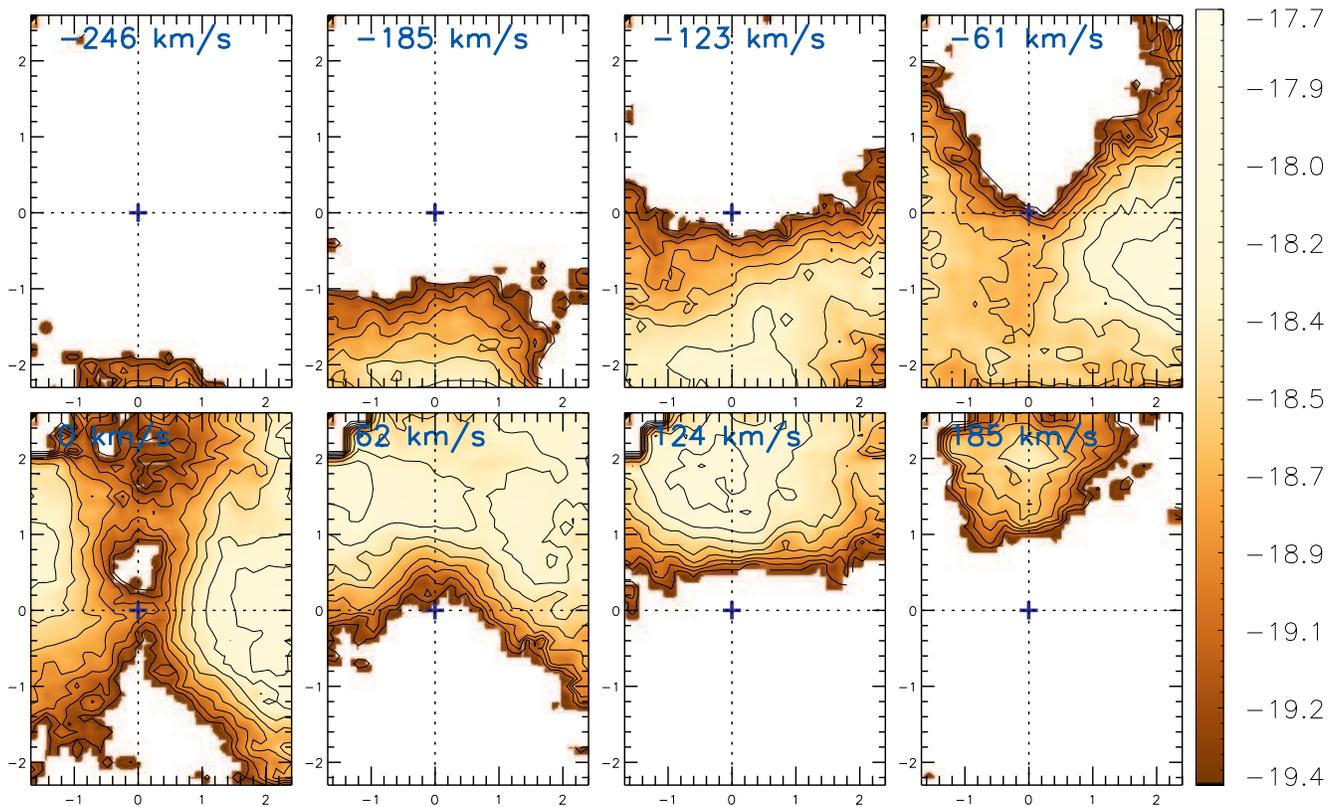}
   \caption{Channel maps of width $\approx$\,60 km\,s$^{-1}$ -- along the emission-line profiles of H$\beta$. The numbers in the top left corners are the central 
            velocities in km\,s$^{-1}$. Fluxes are shown in logarithmic scale and in units of erg\,cm$^{-2}$\,s$^{-1}$. Spatial coordinates are in arcseconds. 
            The cross indicates the position of the nucleus (continuum peak).}
   \label{fig:sliceHb}
\end{minipage}
\end{figure*}

In Figure \ref{fig:sliceHb}, we present the channel maps of the H$\beta$ emission line for 8 velocities, integrated in velocity bins of $\approx$60\,\kms. The 
systemic velocity has been subtracted from the maps. Its value was obtained from the fit of the rotation model and also independently by inspection of the \Hb\
velocity channels. Under the assumption that the kinematics in the velocity channels should be dominated by rotation (as observed in the centroid velocity map)  
we have adopted the value of the systemic velocity as the one for which the blueshifts reach similar values to the redshifts, as this symmetry should be present 
in a rotation pattern. This value is $\approx$3960\,\kms. This \qm{technique} for obtaining the centroid velocity is particularly useful in the case that the 
gas velocity field is disturbed by non-circular motions (e.g., Paper\,I).

In the channel maps of Fig.\ref{fig:sliceHb}, the highest negative velocities reach $-$246\,\kms\ to the West of the nucleus, while the highest positive 
velocities are observed to the East of the nucleus, reaching $\approx$245\,\kms (although this velocity channel is not shown in Figure \ref{fig:sliceHb}). 
These highest values support an orientation for the kinematic major axis along the vertical axis of our FOV. The velocities along the minor axis are close 
to zero, in accordance to what is expected for circular motion.

\begin{figure*}
\begin{minipage}[c]{1.0\textwidth}
\centering 
  \includegraphics[scale=0.90,angle=0]{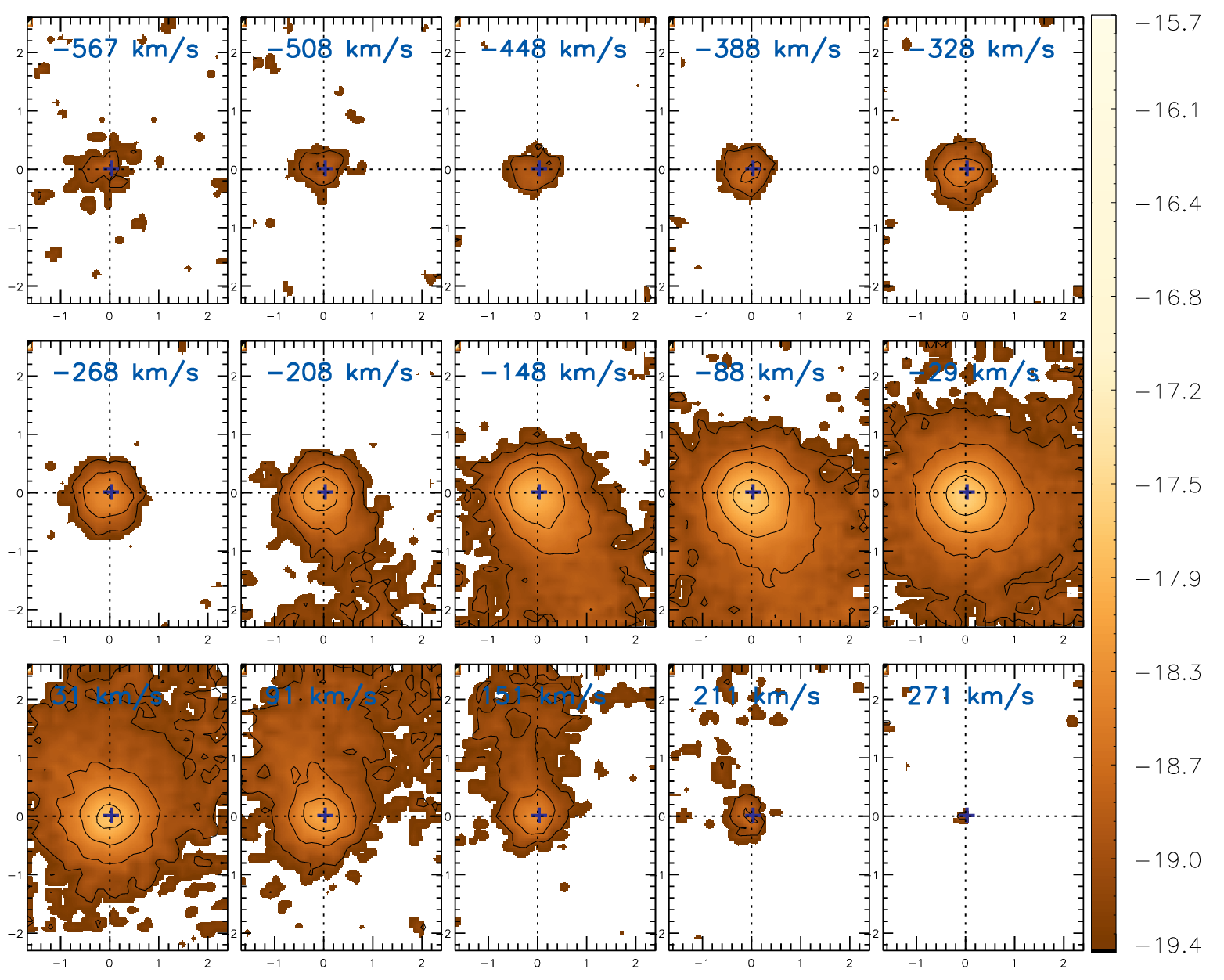}
    \caption{Channel maps of width $\approx$\,60 km\,s$^{-1}$ -- along the emission-line profiles of \oiii. The numbers in the top left corners are the central 
            velocities in km\,s$^{-1}$. Fluxes are shown in logarithmic scale and in units of erg\,cm$^{-2}$\,s$^{-1}$. Spatial coordinates are in arcseconds. 
            The cross indicates the position of the nucleus (continuum peak).}
\label{fig:sliceOIII}
\end{minipage}
\end{figure*}

In Figure \ref{fig:sliceOIII}, we present the channel maps in the \oiii\ emission line. For \oiii\ we show 12 velocity channels, integrated again in velocity 
bins of 60\kms. The highest negative velocities are observed at the nucleus, within a radius of 0\farcs5, and reach $-567$\,km\,s$^{-1}$. The highest positive 
velocities reach only $200$\,\kms. The velocity bins between $-200$ and $200$\,\kms\ seem to be dominated by circular motion.

\section{Discussion}
\label{sec:discussions}

\subsection{Stellar Population}  \label{sec:diss.stelpop}

From the results of the stellar population synthesis we obtain the spatial variations of the contributions of each SPCs, finding that, within the inner 0.26 kpc (radius), 
the old population dominates with up to 85 per cent of the flux at 5100\AA. The young stellar population also contributes to the flux in this central region, while the 
intermediate age stellar population is only observed in the elongated half ring extending from the East through North to the West of the nucleus. Its location in the 
plane of the galaxy is co-spatial with the inner border of the North side of the H$\beta$ ring at $\approx$\,500\,pc from the nucleus.

Beyond the inner 0.26\,kpc, the dominant SPC is the young stellar population (with age $\le 25$\,Myr), contributing with at least 85 per cent of the flux at 5100 \AA. 
Ionizing stars are required to be present in this region for consistency with the corresponding emission-line ratio values in the BPT diagram (Sec.\,\ref{sec:dis.excitation}) 
which indicate that the gas excitation is characteristic of HII regions.

A comparison between the SDSS spectrum and the IFU integrated spectrum (Fig. \ref{fig:sdss+ifu}) shows that both present features usually attributed to a post-starburst 
population. However, when the synthesis is performed with the spatial resolution allowed by the IFU data, the result is that there is no intermediate stellar population 
at the nucleus, only at the circumnuclear ring. We can only conclude that the strong signature of an intermediate age stellar population in the SDSS spectrum is due to 
the contribution of the intermediate age population at the ring combined with the contribution of both the old and young SPCs at the ring, which is partially included 
in the SDSS aperture of 3\arcsec. 

From the gas kinematics, we also conclude that non-circular motions -- in particular, a nuclear outflow -- is observed within the inner 0.26\,kpc radius, where the old stellar
population is dominant and there is no significant intermediate age population contribution. The absence of intermediate age stellar population in this region indicates that 
there has been no quenching of star formation within the inner 0.26\,kpc.

The intermediate age stellar population is observed only in the ring beyond the inner 0.26\,kpc. We cannot exclude the possibility that there could have been a more 
powerful outflow from the AGN in the past that could have extended out to the ring. But in order to argue that such an outflow has quenched the star formation originating 
the post-starburst population now observed, the history of star formation in the ring should reveal a gap between the ages of the post starburst stellar population and 
that of the young stellar population observed there. But our synthesis does not show any significant gap in the contribution of the stellar population with ages ranging 
from 5\,Myr and 900\,Myr, supporting instead continuous star formation in the ring since 900\,Myr ago. This suggests that the post-starburst stellar population in the 
ring is not due to quenching by the AGN.

\subsection{Stellar Kinematics}  \label{sec:dis.stellarkinematics}
 
The stellar velocity field presents a rotation pattern (Fig. \ref{fig:starkinematics}) with the line of nodes running approximately vertically in the figure, showing 
blueshifts to the West and redshifts to the East. The maximum rotation amplitude is $\approx$\,70\,\kms, while the the velocity dispersion values reach $\approx$110\,\kms.

We have estimated the mass of the supermassive black hole (SMBH) in the center of the \psqC\ from the bulge stellar velocity dispersion by using the local $M_\rmn{BH}-\sigma_{*}$
relation of \cite{kormendy13}: $\rmn{log}(\rmn{M}_\rmn{BH}/10^9\rmn{M}_\odot)=(-0.501\pm0.049)+(4.414\pm0.295)\,\rmn{log}(\frac{\sigma_*}{200})$. We have adopted $\sigma_* =
100\pm7$\,\kms\ for the velocity dispersion in the region dominated by the old bulge stellar population, obtaining a mass value for the SMBH of 
$\rmn{log}(\rmn{M}_\rmn{BH}/\rmn{M}_\odot) = 7.17 \pm 0.16$.

The above mass value can be compared to two previous estimates obtained by \cite{greene06} and \cite{shen08}, which are, respectively $\rmn{log}(\rmn{M}_\rmn{BH}/\rmn{M}_\odot) = 7.0 \pm 
0.1$ and $\rmn{log}(\rmn{M}_\rmn{BH}/\rmn{M}_\odot) = 7.38 \pm 0.07$. These latter measurements are based on the SDSS integrated spectrum and the mass-scale relationship of 
\cite{vestergaard06}, using the FWHM of the broad H$\beta$ and the continuum luminosity at 5100\,\AA. Our measurement is based on the stellar velocitydispersion value of the 
galaxy bulge and we thus believe that it is more robust than these previous values. On the other hand, our resulting SMBH mass value is similar to those obtained in these 
previous studies.

\subsection{Gas Excitation}
\label{sec:dis.excitation}

\begin{figure*}
\begin{minipage}[c]{1.0\textwidth}
\centering 
\includegraphics[scale=0.5,angle=0]{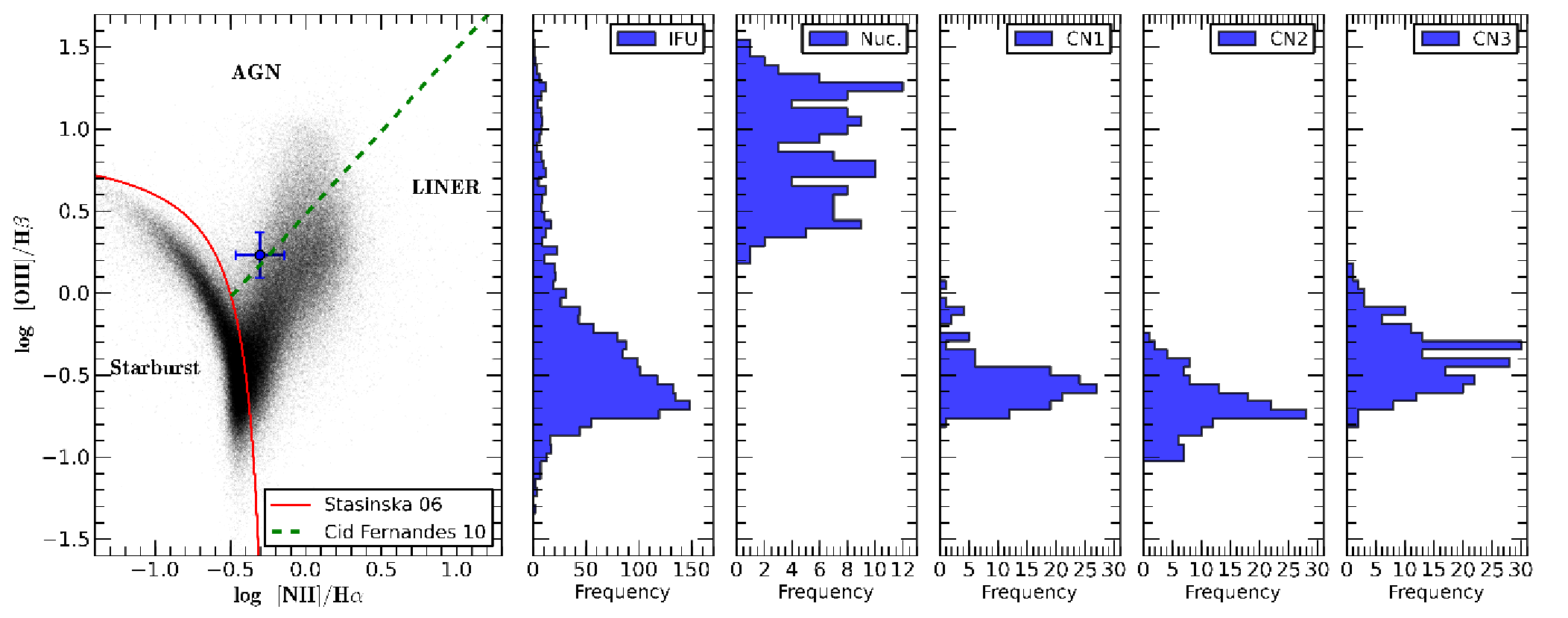}
\caption{Leftmost panel: Baldwin, Phillips \& Terlevich (1981) diagram \oiii/\Hb\ $\times$ \nii/\ha\ showing the locus of \psqC\ (shown as a filled blue circle 
         with its error bars) obtained from the line ratios of the SDSS spectrum. Histograms, from left to right: the first shows the ratios from individual 
         spectra within 3\farcs0 from the nucleus (the same aperture of the SDSS spectrum). The second shows the ratios for the region labeled as Nuc. The next three show the 
         ratios from the circumnuclear regions CN1, CN2 and CN3 in Fig. \ref{fig:flux}.}
\label{fig:bpt}
\end{minipage}
\end{figure*}

In the leftmost panel of Figure \ref{fig:bpt} we show the \cite{bpt81} diagram \oiii/\Hb\ $\times$ \nii/\ha\ with the position of \psqC\ (shown as a filled blue circle with 
its error bars) as derived from the line ratios in the SDSS spectrum. The locus in the diagram is just to the right of the limiting region between the loci of Starburst and 
Seyfert galaxies, drawn as a solid line in the diagram (from \cite{stasinska06}). The dashed line -- limiting line between the Seyfert AGNs and LINERs -- is from \cite{cid10}.

Since the SDSS spectrum aperture covers almost the whole field-of-view (FOV) of our observations, we have used our data in order to map the emission line ratios 
within our FOV, not only at the nucleus, but also in the circumnuclear region. But as our IFU spectra do not cover the H$\alpha$ region, we decided to show the 
\oiii/\Hb\ emission line ratio values as histograms, assuming a fixed \nii/\ha\ equal to that of the SDSS spectrum. We also show these histograms in Fig.\,
\ref{fig:bpt} for the different regions identified in Figure \ref{fig:flux}.

The first histogram, labeled as IFU, shows the ratios for all spectra within 1\farcs5 from the nucleus (the same aperture of the SDSS spectrum). The second histogram 
shows the ratios for the spectra located inside the region labeled as Nuc. The other three histograms (labeled CN1, CN2 and CN3) show the emission line ratios from the 
spectra of three circumnuclear regions located in the 500\,pc H$\beta$ ring. The corresponding line ratios are typical of HII regions, while the line ratios from the 
nucleus are typical of Seyfert nuclei.  

\subsection{Mass of emitting gas}
\label{sec:mass}

We have estimated the total mass of ionized gas within the inner 1\farcs5 ($\approx$400\,pc) radius as:
\begin{equation}
M = V \epsilon\,n_{\rm e} m_{\rm p},
\label{eq:mass0}
\end{equation}
where $n_{\rm e}$ is the electron density, $m_{\rm p}$ is the proton mass, $\epsilon$ is the filling factor and $V$ is the volume of the emitting region.

Following Paper\,I and \cite{peterson97}, we have estimated the product $V\epsilon$ by:
\begin{equation}
V\epsilon = 8.1 \times 10^{59}\dfrac{L_{41}({\rm H}\beta)}{{n_3}^2}\,\,\,\rmn{cm}^{-3},
\label{eq:epsilon}
\end{equation}
where $L_{\rm 41}({\rm H\beta})$ is the H$\beta$ luminosity, in units of $10^{41}$\,ergs\,s$^{-1}$ and $n_3$ is the electron density in units
of $10^3$\,cm$^{-3}$. The mass of the emitting region can be obtained by using:
\begin{equation}
M\approx7\times10^{5}\dfrac{L_{\rm 41}({\rm H\beta})}{n_3}\,\,\,M_\odot\,,
\label{eq:mass1}
\end{equation}
given in units of solar masses.

The H$\beta$ luminosity was calculated from the integrated H$\beta$ flux F(H$\beta$) of the inner 1\farcs5 region corrected for the reddening obtained from the SDSS spectrum
($C({\rm H\beta}) = 1.51 \pm 0.12$), using the reddening law of \cite{cardelli89} adopting the theoretical $\frac{F(\rmn{H}\alpha)}{F(\rmn{H}\beta)}$ ratio of 3.0 (case B 
recombination of \citet{osterbrock06}).  For the assumed distance of $d=53.5 \pm 3.7$\,Mpc, we obtain $L({\rm H\beta})=4 \upi d^2 F({\rm H}\beta)10^{C({\rm H}\beta)} = 2.32 
\pm 0.53 \times 10^{40}$\,ergs\,s$^{-1}$.
 
The electron density $n_3$ was obtained using the [S\,{\scriptsize II}] \,$\lambda6716/\lambda6731$ ratio of the SDSS spectrum, by solving numerically the equilibrium equation 
for a $5$-level atom using the {\small IRAF} routine \textsc{stsdas.analysis.nebular.temden} \citep{derobertis85,shaw94}. For an assumed electron temperature of $16000$\,K 
\citep{peterson97}, we obtain $n_\rmn{e}=263^{+52}_{-47}$\,cm$^{-3}$. Using the above derived values for $L({\rm H\beta})$ and  $n_\rmn{e}$ we obtain a mass of ionized gas 
within the inner 400\,pc radius of $6.2 \pm 1.4 \times 10^5 M_\odot$. This value is about 0.1 of that obtained in Paper\,I and for other Seyfert galaxies \citep{sb09}.

\subsection{Gas Kinematics}
\label{sec:diss.kinematics}

The velocity field of the H$\beta$ emitting gas presented in Fig. \ref{fig:cinematica} shows a rotation pattern with apparently similar rotation axis and kinematic center 
to that of the stellar velocity field. However, the velocity amplitude of the gas kinematics reaches $\approx$130\,\kms, while the amplitude in the stellar velocity reach 
only $\approx$70\,\kms, what can be attributed to the effect of asymmetric drift \citep{verdoes00,barth01}. The H$\beta$ channel maps (Fig.\ref{fig:sliceHb}) confirm the dominance 
of a rotation pattern in the H$\beta$ kinematics.
 
In the case of \oiii, both the centroid velocity field (Fig. \ref{fig:cinematica}) and channel maps (Fig.\ref{fig:sliceOIII}) show the presence of a blueshifted component 
besides rotation. In the centroid velocity field, blueshifts are observed at the nucleus instead of zero velocity (as is the case of the H$\beta$ velocity field). In the 
channel maps there is \oiii\ emission within 0\farcs5 (130\,pc) from the nucleus in many blueshifted channels -- from negative velocities as high as $-560$\,\kms\ down to 
$-200$\,\kms. For lower velocities, the rotation component dominates. There is no high velocity nuclear component in redshift.
 
\begin{figure*}
\begin{minipage}[c]{1.0\textwidth}
\centering 
\includegraphics[scale=0.7, angle=0]{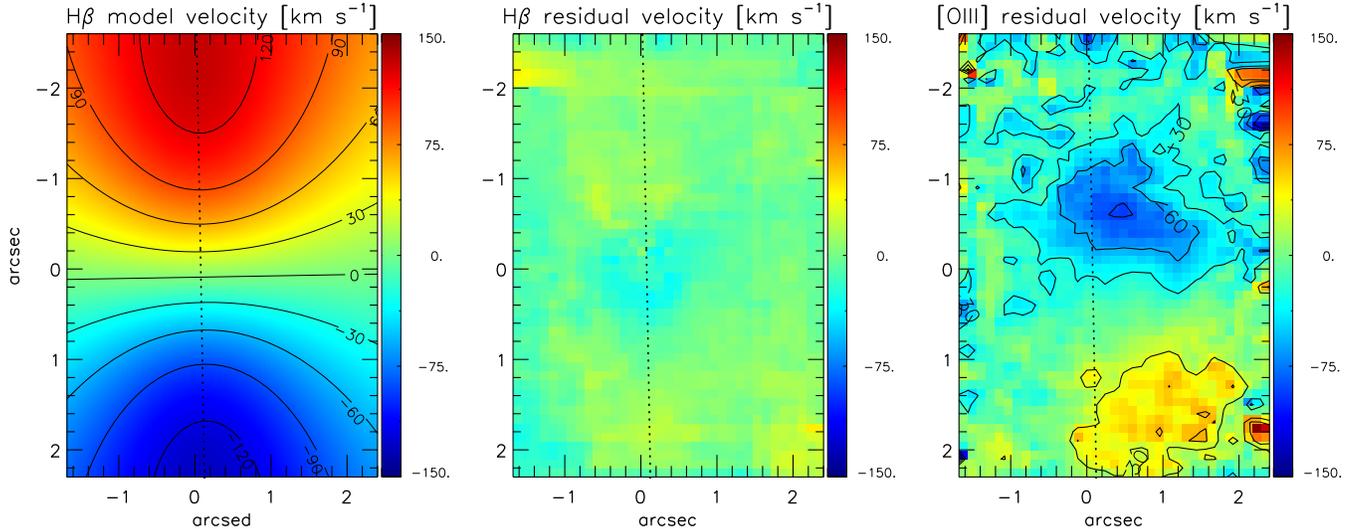}
 \caption{Rotation model for the gas kinematics (left panel), the \Hb\ residual velocity map (central panel) and the \oiii\ residual velocity map (right panel). 
          The residual velocity maps were obtained from the subtraction of the model from the observed velocities.}
 \label{fig:modeloGas}
\end{minipage}
\end{figure*}

\subsubsection{Rotation model} \label{sec:rotation}

Since the gas kinematics of the \Hb\ emitting gas seems to present a clear rotation pattern, we have fitted a rotation model to the H$\beta$ velocity field. 
This velocity field will also be used to represent the rotation component in the \oiii\ velocity field, which clearly shows other kinematic components, and its 
subtraction will allow us to isolate these non-circular components.

Following \cite{bertola91} we have approximated the gas rotation by Keplerian motion in a circular disk in a spherical potential, for which the rotation curve is 
given by 
\begin{equation}
v_c(r) = \frac{A\,r}{\paren{r^2+c_0^2}^{p/2}} \label{eq:vc}\,,
\end{equation}
where $A$, $c_0$ and $p$ are the parameters and $r$ is the radius in the plane of the galaxy. As given in \cite{bertola91}, the observed radial velocity projected 
at a position $(R,\Psi)$ in the plane of the sky is:
\begin{equation}
\begin{split}
v_\rmn{}(R,\Psi) = v_\rmn{sys} + \hspace{5.2cm} \\ 
+ \frac{AR\,\cos{(\Psi-\Psi_0)}\,\sin\theta\,\cos^p\theta}{\{ {R^2\squares{\sin^2(\Psi-\Psi_0)+\cos^2\theta 
\cos^2(\Psi-\Psi_0)}+C_0^2 \}}^{p/2}} \label{eq:vmod}\,,
\end{split}
\end{equation}
where $R$ is the projected distance on the plane of the sky relative to the kinematics center $(X_0,Y_0)$, $\Psi$ is the corresponding position angle of $R$ relative 
to the position angle of the line of the nodes ($\Psi_0$), $v_\rmn{sys}$ is the systemic velocity, $\theta$ is the inclination of the disc (with $\theta=0$ for a 
face-on disk) and $C_0^2=c_0^2 \cos^2\theta$. The model has a total of $8$ free parameters ($v_\rmn{sys}$, $A$, $c_0$, $p$, $\Psi_0$, $\theta$, $X_0$ and $Y_0$), 
but we have fixed the kinematic center ($X_0$ and $Y_0$), adopted as the position of the peak flux in the continuum, the position angle of the line of the nodes, 
adopted as the photometric major axis at $PA=89°$ \citep{graham09} and the inclination $\theta=53°$ \citep{graham09} of the disc of the galaxy. 

The other parameters were determined by fitting the model to the observed \Hb\ velocity field via a Levenberg-Marquardt least squares ﬁtting algorithm. Initial guesses
were given for the free parameters and their errors were estimated directly from the fitting algorithm for an adopted mean uncertainty of 10\,\kms. 
From the fit we obtain a systemic velocity of $v_\rmn{sys}=3965\pm5$\,\kms, and the parameter values $A=663\pm142$, $c_0=0.60\pm0.05$\,kpc and $p=1.89\pm0.13$.

In Figure \ref{fig:modeloGas} we show the best-fit rotation model (left panel) for the \Hb\ kinematics, the \Hb\ residual velocity map (central panel) --  
obtained as the difference between the observed \Hb\ velocities and the modeled velocities, and the \oiii\ residual velocity map (right panel) -- also obtained 
as the difference between the observed \oiii\ and the modeled \Hb\ velocities. Most \Hb\ residual velocity values are smaller than $\approx$25\,\kms, which represents about 
20 per cent of the velocity amplitude of the rotation model ($v_\rmn{max} \approx 130$\,\kms). 

In the right panel of Figure \ref{fig:modeloGas}, we show the \oiii\ residual map. Two regions show significant residuals, one with residual blueshifts and the other 
with redshifts. Blueshifts are observed from the nucleus up to 170\,pc from it towards the Northeast, with velocity values reaching up to $\approx$-80\,\kms. Redshift 
are observed at a similar distance to the Northwest and reach values of up to $\approx$40\,\kms. 

\subsection{Mass outflow rate}
\label{sec:outflowrate}

As in Paper\,I, we assume that the outflow in \psqC\ has a conical geometry, with the cone axis directed approximately towards us. We thus consider that the outflowing gas 
is crossing the base of this cone. The radius of this circular base is estimated from the extent of the blueshifted region seen in the channel maps, of about 0\farcs5. 
We have thus measured the ionized gas mass outflow rate at negative velocities through a circular cross section with radius $r = 0.13$\,kpc (0\farcs5) around the nucleus, 
assuming that the height of the cone $h$ is equal to the diameter of its base. 

We have calculated the mass outflow rate using the method described in \cite{canodiaz12}. Assuming the geometry described above and considering an outflow at a distance 
$h_\rmn{kpc}$ (in units of kpc) from the nucleus, the outflow rate in ionized gas can be obtained from:
\begin{equation}
\dot{M}_\rmn{out}=164\,\frac{L_\rmn{44}(\rmn{\oiii})\,v_\rmn{3}}{n_\rmn{3}\,10^{\rmn{[O/H]}} h_\rmn{kpc}}\, M_\odot\,\rmn{yr}^{-1}\,,
\end{equation}
where $L_\rmn{44}$(\oiii) is the luminosity (in units of 10$^{44}$\,erg\,s$^{−1}$) of the \oiii\ emission line tracing the outflow, $n_\rmn{3}$ is the electron 
density (in units of 10$^3$ cm$^{-3}$) and 10$^\rmn{[O/H]}$ is the oxygen abundance in solar units, which we have assumed to be one. With these assumptions, we obtain 
a mass outflow rate of $\dot{M}_\rmn{out} \approx 0.03\, \rmn{M_\odot}$\,yr$^{-1}$.

In order to compare the mass outflow rate with the accretion rate necessary to feed the AGN ($\dot{m}$), we have used the following equation \citep{peterson97}:
\begin{equation}
\dot{m}=\dfrac{L_\rmn{bol}}{c^2\,\eta} \approx 1.8 \times 10^{-3} \left( \dfrac{L_{44}}{\eta}\right) \,\,{\rm M_\odot}\,\rm{yr^{-1}} \,,
\label{eq:mdot}
\end{equation}
where $L_\rmn{bol}$ is the nuclear bolometric luminosity, $c$ is the light speed and $\eta$ is the efficiency of conversion of the rest mass energy of the accreted material 
into radiation power and $L_{44}$ is bolometric luminosity in units of 10$^{44}$\,erg\,s$^{-1}$. As in our previous paper (Paper\,I) and following \cite{rogemar11a,rogemar11b}, 
$L_\rmn{bol}$ was approximated by$\approx\,100\,L(\rmn{H}\alpha)$, where $L(\rmn{H}\alpha)$ is the H$\alpha$ nuclear luminosity. Using the interstellar extinction coefficient 
$C(\rmn{H}\beta)$, the intrinsic ratio $\frac{F(\rmn{H}\alpha)}{F(\rmn{H}\alpha)}=3.0$ \citep{osterbrock06} and the reddening law of \cite{cardelli89}, we obtain 
$F(\rmn{H}\alpha) \approx 1.51 
\times 10^{-13}$\,erg\,cm$^{-2}$\,s$^{-1}$ within 0\farcs3 of the nucleus. At the distance of $\approx 53$\,Mpc, we have $L(\rmn{H}\alpha) \approx 5.20 \times 10^{40}$\,erg\,s$^{-1}$,
implying that the corresponding nuclear bolometric luminosity is $L_\rmn{bol} \approx 5.20 \times 10^{42}$\,erg\,s$^{-1}$. Assuming an efficiency $\eta \approx 0.1$, for an 
optically thick and geometrically thin accretion disc \citep{frank02} we derive an accretion rate of $\dot{m} = 9.3 \times 10^{-4}\,\rmn{M_\odot}$\,yr$^{-1}$.

Comparing our estimate of mass outflow rate with the accretion rate $\dot{m}$, we note that $\dot{M}_\rmn{out}$ is approximately one order of magnitude higher than 
$\dot{m}$. This implies that the outflowing gas can not originate from the AGN, but is rather mostly composed of interstellar gas from the surrounding region 
of the galaxy being pushed away from the central region by a nuclear outflow. 
 
We can additionally estimate the kinetic power of the outflow, considering both the radial and turbulent component of the gas motion, as follows:
\begin{equation}
\dot{E}_\rmn{out} \approx \frac{\dot{M}_\rmn{out}}{2}(v_\rmn{out}^2+\sigma^2)\,,
\label{eq:power}
\end{equation}
where $v_\rmn{out}$ is the velocity of the outflowing gas and $\sigma$ is the velocity dispersion. 
From Fig. \ref{fig:cinematica} we have $\sigma \approx 100$ \,km\,s$^{-1}$ and adopting the $v_\rmn{out}=221$\,km\,s$^{-1}$ estimated above, we obtain
that $\dot{E}_\rmn{out} = 5.6 \times 10^{38} $erg\,s$^{-1}$, which is $\approx 1.1 \times 10^{-4} L_\rmn{bol}$.

\section{Summary and Conclusions}
\label{sec:conclusions}

We have used integral field optical spectroscopy (GMOS-IFU) in order to map the stellar population and the kinematics in the central kiloparsec of the Post-Starburst 
Quasar \psqC\ at a spatial resolution of $\approx$\,130\,pc (0\farcs5) and velocity resolution of $\approx$40\,\kms. This is the first two-dimensional study 
of the stellar population and both the gas and stellar kinematics of a Post-Starburst Quasar. Our main conclusions are:

\begin{enumerate}

\item The stellar population is dominated by the old age component (SPC)  within the inner 260\,pc radius, while in the circumnuclear region (at $\approx$350\,pc 
from the nucleus) the young SPC ($t<100$\,Myrs) dominates the optical flux. The post-starburst component (100\,Myrs $< t \le$ 2.5G\,Gyrs) is exclusively distributed 
along a half-ring at $\approx$500\,pc North from the nucleus;
 
\item Both \oiii\ and \Hb\ emission are extended all over the mapped field. The \oiii\ flux distribution is brighter in the nucleus, while the \Hb\ flux is brighter
along an almost complete ring at $\approx$0.5\,kpc from the nucleus; 

\item The emission line ratios in the circumnuclear ring are typical of HII regions, implying that the gas in this region is ionized by young stars, what is supported 
by the dominant age component at this location ($\leq$25\,Myrs). In the inner 260\,pc the emission line ratios are typical of Seyfert excitation;

\item The mass of ionized gas in the inner 1\farcs5 radius (400\,kpc) is $6.2 \pm 1.4 \times 10^5 M_\odot$;

\item The \Hb\ gas kinematics was reproduced by a model of rotation in the plane of the galaxy, for which the line of nodes runs approximately along the East-West 
direction and has a velocity amplitude of $\approx130$\,km\,s$^{-1}$;

\item The \oiii\ kinematics, besides rotation, also shows an outflow reaching $\approx -500$\,\kms\ within $\approx$250\,pc of the nucleus;
 
\item We have estimated a mass outflow rate of $\approx$0.03\,M$_\odot$\,yr$^{-1}$, that is $\approx$ 30 times the AGN mass accretion rate 
($\approx 9.3 \times 10^{-4}\,\rmn{M_\odot}$\,yr$^{-1}$). This implies that most of the outflow originates in mass-loading of a nuclear outflow 
as it moves outwards pushing the interstellar medium of the galaxy in the vicinity of the nucleus.

\item From the stellar velocity dispersion we have estimated a mass for the SMBH of $\rmn{log}(\rmn{M}_\rmn{BH}) = 7.17 \pm 0.16$.

\end{enumerate}

Regarding the favored scenario for the origin of the post-starburst population, our study does not support the quenching scenario, as the post-starburst population
is not located close enough to the nucleus, where the outflow is observed. It is instead located in a ring at $\approx$500\,pc from the nucleus, which is out of the 
reach of the AGN outflow (feedback). Although we cannot exclude the possibility that there could have been a more powerful outflow from the AGN in the past that could 
have quenched star formation in the ring, the history of star formation supports instead continuous star formation since 900\,Myr ago. This suggests that the post-starburst 
stellar population in the ring is not due to quenching by the AGN. In the more central region, internal to the ring, where we now observe the outflow, the stellar population 
is old with some contribution from young stars and does not show any signature of star formation quenching as well. In Paper\,I we found that both scenarios could be at play: 
quenching due to the AGN feedback close to the nucleus (as we found a post-starburst population within $\approx$300\,pc from the nucleus), co-spatial with the nuclear outflow, 
and the evolutionary scenario for the ring at $\approx$800\,pc from the nucleus. Our goal now is to extend this kind of analysis to similar objects using integral field 
spectroscopy in order to verify how common are these scenarios and conclude which is the main mechanism responsible for the presence of a post-starburst population 
in PSQs and AGNs in general.

\section*{Acknowledgments}

Based on observations obtained at the Gemini Observatory, which is operated by the Association of Universities for Research in Astronomy, Inc., under a cooperative 
agreement with the NSF on behalf of the Gemini partnership: the National Science Foundation (United States), the Science and Technology Facilities Council (United Kingdom), 
the National Research Council (Canada), CONICYT (Chile), the Australian Research Council (Australia), Minist\'erio da Ci\^ncia, Tecnologia e Inova\c{c}\~ao (Brazil) and 
Ministerio de Ciencia, Tecnolog\'ia e Innovaci\'on Productiva (Argentina). This research has made use of the NASA/IPAC Extragalactic Database (NED) which is operated by 
the Jet Propulsion Laboratory, California Institute of Technology, under contract with the National Aeronautics and Space Administration. The {\small STARLIGHT} project 
is supported by the Brazilian agencies CNPq, CAPES and FAPESP, and by the France-Brazil CAPES/Cofecub program. This work has been partially supported by the Brazilian 
institution CNPq.

\bsp

\label{lastpage}

\begin{thebibliography}{99}

\bibitem[\protect\citeauthoryear{Baldwin, Phillips \& Terlevich}{1981}]{bpt81} {Baldwin} J.~A., {Phillips} M.~M., {Terlevich} R., 1981, PASP, 93, 5

\bibitem[\protect\citeauthoryear{Barth et al.}{2001}]{barth01} {Barth} A.~J., {Sarzi} M., {Rix} H.~W., {Ho} L.~C., {Filippenko} A.~V., {Sargent}, W.~L.~W., 2001, ApJ, 555, 685

\bibitem[\protect\citeauthoryear{Bertola et al.}{1991}]{bertola91} {Bertola} F., {Bettoni} D., {Danziger} J., {Sadler} E., {Sparke} L., {de Zeeuw} T., 1991, ApJ, 373, 369

\bibitem[\protect\citeauthoryear{Bruzual \& Charlot}{2003}]{bc03} {Bruzual} G., {Charlot} S., 2003, MNRAS, 344, 1000

\bibitem[\protect\citeauthoryear{Brotherton et al.}{1999}]{brotherton99} {Brotherton} M.S., {van Breugel} W., {Stanford} S.A., {Smith} R.J.,
{Boyle} B.J., {Miller} L., {Shanks} T., {Croom} S.M., {Filippenko} A.V., 1999, ApJL, 520, 87

\bibitem[\protect\citeauthoryear{Cales et al.}{2011}]{cales11} {Cales} S.L., {Brotherton} M.~S., {Shang} Z., {Bennert} V.~N., {Canalizo} G., {Stoll} R., {Ganguly} R.,
{Vanden Berk} D., {Paul} C., {Diamond-Stanic} A., 2011, ApJ, 741, 106 

\bibitem[\protect\citeauthoryear{Cales et al.}{2013}]{cales13} {Cales} S.~L., {Brotherton} M.~S., {Shang} Z., {Runnoe} J.~C., {DiPompeo} M.~A., {Bennert} V.~N., {Canalizo} G.,
{Hiner} K.~D., {Stoll} R., {Ganguly} R., {Diamond-Stanic} A., 2013, ApJ, 762, 90 

\bibitem[\protect\citeauthoryear{Cano-D{\'{\i}}az et al.}{2012}]{canodiaz12} {Cano-D{\'{\i}}az} M., {Maiolino} R., {Marconi} A., {Netzer} H., {Shemmer} O., 
{Cresci} G., 2012, A\&A, 537, 8

\bibitem[\protect\citeauthoryear{Cappellari \& Emsellem}{2004}]{cappellari04} Cappellari M. and Emsellem E., 2004, PASP, 116, 138

\bibitem[\protect\citeauthoryear{Cardelli, Clayton \& Mathis}{1989}]{cardelli89} Cardelli J. A., Clayton G.C., Mathis J.S., 1989, ApJ, 345, 245

\bibitem[\protect\citeauthoryear{Chabrier}{2003}]{chabrier03} {Chabrier} G., 2003, PASP, 115, 763 

\bibitem[\protect\citeauthoryear{Cid Fernandes et al.}{2001}]{cid01} {Cid Fernandes} R., {Sodr{\'e}} L., {Schmitt} H.~R., {Le{\~a}o} J.~R.~S., 2001, MNRAS, 325, 60

\bibitem[\protect\citeauthoryear{Cid Fernandes et al.}{2004}]{cid04} Cid Fernandes R., Gu, Q. Melnick J., Terlevich E., Terlevich R., Kunth D., 
Rodrigues L.R., Joguet B., 2004, MNRAS, 355, 273

\bibitem[\protect\citeauthoryear{Cid Fernandes et al.}{2005}]{cid05} Cid Fernandes R., Mateus A., Sodr\'e L., Stasi\'nska G., Gomes J.M., 2005, MNRAS, 358, 363

\bibitem[\protect\citeauthoryear{Cid Fernandes et al.}{2009}]{cid09} {Cid Fernandes} R., {Schoenell} W., {Gomes} J.~M., {Asari} N.~V., {Schlickmann} M., {Mateus} A., {Stasi{\'n}ska} G.,
{Sodr{\'e}} Jr.~L., {Torres-Papaqui} J.~P. and {Seagal Collaboration}, 2009, RMXAC, 35, 127

\bibitem[\protect\citeauthoryear{Cid Fernandes et al.}{2010}]{cid10} {Cid Fernandes} R., {Stasi{\'n}ska} G., {Schlickmann} M.~S., {Mateus} A., {Vale Asari} N., {Schoenell} W.,
{Sodr{\'e}} L., 2010, MNRAS, 403, 1036

\bibitem[\protect\citeauthoryear{Davies et al.}{2007}]{davies07} {Davies} R.~I., {M{\"u}ller S{\'a}nchez} F., {Genzel} R., {Tacconi} L.~J., {Hicks} E.~K.~S., 
{Friedrich} S., {Sternberg} A., 2007, ApJ, 671, 1388

\bibitem[\protect\citeauthoryear{Di Matteo et al.}{2005}]{dimatteo05} {Di Matteo} T., {Springel} V., {Hernquist} L., 2005, Nat, 433, 604

\bibitem[\protect\citeauthoryear{De Robertis et al.}{1987}]{derobertis85} {De Robertis} M.M., {Dufour} R.J., {Hunt} R.W., 1987, JRASC, 81, 195

\bibitem[\protect\citeauthoryear{Frank, King \& Raine}{2002}]{frank02} {Frank} J., {King} A., {Raine}, D.J., 2002, Accretion Power in Astrophysics, 3rd ed.,
Cambridge Univ. Press, Cambridge

\bibitem[\protect\citeauthoryear{Graham \& Li}{2009}]{graham09} Graham A.W.,  Li I-hui., 2009, ApJ, 698, 812

\bibitem[\protect\citeauthoryear{Graham et al.}{2011}]{graham11} {Graham} A.~W., {Onken} C.~A., {Athanassoula} E., {Combes} F., 2011, MNRAS, 412, 2211

\bibitem[\protect\citeauthoryear{Granato et al.}{2004}]{granato04} {Granato} G.~L., {De Zotti} G., {Silva} L., {Bressan} A., {Danese} L., 2004, ApJ, 600, 580

\bibitem[\protect\citeauthoryear{Greene \& Ho}{2006}]{greene06} {Greene} J.~E., {Ho} L.~C., 2006, ApJ, 641, 21


\bibitem[\protect\citeauthoryear{Heckman et al.}{2004}]{heckman04} {Heckman} T.M., {Kauffmann} G., {Brinchmann} J., {Charlot} S., {Tremonti} C., 
{White} S.~D.~M., 2004, 613, 109

\bibitem[\protect\citeauthoryear{Ho et al.}{2000}]{Ho00} {Ho} L.~C., {Rudnick} G., {Rix} H.~W., {Shields} J.~C., {McIntosh} D.~H., {Filippenko} A.~V., {Sargent} W.~L.~W.,
{Eracleous} M., 2000, 541, 120

\bibitem[\protect\citeauthoryear{Hopkins et al.}{2006}]{hopkins06} {Hopkins} P.~F., {Hernquist} L., {Cox} T.~J., {Robertson} B., {Springel} V., 2006, ApJS, 163, 50

\bibitem[\protect\citeauthoryear{Kormendy \& Ho}{2013}]{kormendy13} {Kormendy} J., {Ho} L.~C., 2013, ARA\&A, 51, 511

\bibitem[\protect\citeauthoryear{Le Borgne et al.}{2003}]{leborgne03} {Le Borgne} J.F., {Bruzual} G., {Pell{\'o}} R., {Lan{\c c}on} A., {Rocca-Volmerange} B.,
{Sanahuja} B., {Schaerer} D., {Soubiran} C., {V{\'{\i}}lchez-G{\'o}mez} R., 2003, A\&A, 402, 433

\bibitem[\protect\citeauthoryear{Markwardt}{2009}]{markwardt09} Markwardt, C.B., 2009, ASPCS, 411, 251

\bibitem[\protect\citeauthoryear{Osterbrock \& Ferland}{2006}]{osterbrock06} {Osterbrock} D.E. and {Ferland} G.J., 2006,
Astrophysics of Gaseous Nebulae and Active Galactic Nuclei, 2nd. ed., University Science Books, California

\bibitem[\protect\citeauthoryear{Peterson}{1997}]{peterson97} {Peterson} B.M., 1997, An Introduction to Active Galactic Nuclei, Cambridge Univ. Press, Cambridge

\bibitem[\protect\citeauthoryear{Riffel \& Storchi-Bergmann}{2011a}]{rogemar11a} Riffel R.A., Storchi-Bergmann T., 2011, MNRAS, 411, 469

\bibitem[\protect\citeauthoryear{Riffel \& Storchi-Bergmann}{2011b}]{rogemar11b} Riffel R.A., Storchi-Bergmann T., 2011, MNRAS, 417, 275

\bibitem[\protect\citeauthoryear{Sanmartim, Storchi-Bergmann \& Brotherton}{2013}]{sanmartim13} {Sanmartim} D., {Storchi-Bergmann} T., {Brotherton} M.~S., 2013, MNRAS, 428, 867 

\bibitem[\protect\citeauthoryear{Schlegel, Finkbeiner \& Davis}{1998}]{schlegel98} {Schlegel} D.~J., and {Finkbeiner} D.~P., {Davis}, M., 1998, ApJ, 500, 525  

\bibitem[\protect\citeauthoryear{Shen et al.}{2008}]{shen08} {Shen} J., {Vanden Berk} D.~E., {Schneider} D.~P., {Hall} P.~B., 2008, AJ, 135, 928

\bibitem[\protect\citeauthoryear{Stasi{\'n}ska et al.}{2006}]{stasinska06} {Stasi{\'n}ska} G., {Cid Fernandes} R., {Mateus} A., {Sodr{\'e}} L., {Asari} N.~V., MNRAS, 
2006, 371, 972

\bibitem[\protect\citeauthoryear{Steiner et al.}{2009}]{steiner09} Steiner J.E., Menezes R.B., Ricci, T.V., Oliveira, A.S., 2009, MNRAS, 395, 64

\bibitem[\protect\citeauthoryear{Shaw \& Dufour}{1994}]{shaw94} {Shaw} R.A. and {Dufour} R.J., 1994, ASPC Ser. 61: Astronomical Data
Analysis Software and Systems III, 61, 327 

\bibitem[\protect\citeauthoryear{Storchi-Bergmann et al.}{1997}]{thaisa97} {Storchi-Bergmann} T., {Eracleous} M., {Ruiz} M.~T., {Livio} M., {Wilson} A.~S., 
{Filippenko} A.~V., 1997, ApJ, 489, 87

\bibitem[\protect\citeauthoryear{Storchi-Bergmann et al.}{2001}]{sb01} {Storchi-Bergmann} T., {Gonz{\'a}lez Delgado} R.M., {Schmitt}, H.R., {Cid Fernandes} R.,
{Heckman} T., 2001, ApJ, 559, 147

\bibitem[\protect\citeauthoryear{Storchi-Bergmann et al.}{2009}]{sb09} {Storchi-Bergmann} T., {McGregor} P.~J., {Riffel} R.~A., {Sim{\~o}es Lopes} R., {Beck} T.,
{Dopita} M., 2009, MNRAS, 394, 1148

\bibitem[\protect\citeauthoryear{Storchi-Bergmann et al.}{2010}]{sb10} {Storchi-Bergmann} T., {Lopes} R.D.S., {McGregor} P.J., {Riffel} R.A., {Beck} T., {Martini} P.,
2010, MNRAS, 402, 819

\bibitem[\protect\citeauthoryear{Strateva et al.}{2006}]{strateva06} {Strateva} I.~V., {Brandt} W.~N., {Eracleous} M., {Schneider} D.~P.,  {Chartas}, G., 2006, ApJ, 651, 749

\bibitem[\protect\citeauthoryear{Strateva et al.}{2008}]{strateva08} {Strateva} I.~V., {Brandt} W.~N., {Eracleous} M., {Garmire} G., 2008, ApJ, 687, 869

\bibitem[\protect\citeauthoryear{van der Marel \& Franx}{1993}]{marel93} van der Marel R.P., Franx M., 1993, ApJ, 407, 525

\bibitem[\protect\citeauthoryear{van Dokkum}{2001}]{vandokkum01} van Dokkum, P. G., 2001, PASP, 113, 1420

\bibitem[\protect\citeauthoryear{Veilleux et al.}{2005}]{veilleux05} {Veilleux} S., {Cecil} G., {Bland-Hawthorn} J., 2005, AR\&A, 43, 769

\bibitem[\protect\citeauthoryear{Verdoes et al.}{2000}]{verdoes00} {Verdoes Kleijn} G.~A., {van der Marel} R.~P., {Carollo} C.~M., {de Zeeuw} P.~T., 2000, AJ, 120, 1221

\bibitem[\protect\citeauthoryear{Vestergaard \& Peterson}{2006}]{vestergaard06} {Vestergaard} M., {Peterson} B.~M., 2006, ApJ, 641, 689

\bibitem[\protect\citeauthoryear{Wei et al.}{2013}]{wei13} {Wei} P., {Shang} Z., {Brotherton} M.~S., {Cales} S.~L., {Hines} D.~C., {Dale} D.~A., {Ganguly} R., {Canalizo} G.,
2013, ApJ, 772, 28

\end{thebibliography}
\end{document}